\documentclass[%
 reprint,
 amsmath,amssymb,
 aps,
]{revtex4-2}

\usepackage{graphicx}
\usepackage{dcolumn}
\usepackage{bm}
\usepackage{xcolor}
\usepackage{gensymb}
\usepackage[normalem]{ulem}
\usepackage{url}

\begin{document}


\title{Straight Lightning as a Signature of Macroscopic Dark Matter}

\author{Nathaniel Starkman, Harrison Winch}%
  \email{nathaniel.starkman@mail.utoronto.ca,\newline harrison.winch@mail.utoronto.ca}
\affiliation{%
 David A. Dunlap Department of Astronomy and Astrophysics / Dunlap Institute, University of Toronto, CA\\
}%

\author{Jagjit Singh Sidhu, Glenn Starkman}
\email{jxs1325@case.edu, gds6@case.edu}
\affiliation{Physics Department/CERCA/ISO Case Western Reserve University
Cleveland, Ohio 44106-7079, USA}

\date{\today \quad Accepted: February 19, 2021}

\preprint{APS/123-QED}

\begin{abstract}

    Macroscopic dark matter (macros) is a broad class of alternative candidates to particle dark matter. These candidates would transfer energy to matter primarily through elastic scattering. A sufficiently large macro passing through the atmosphere would produce a straight channel of ionized plasma. If the cross-section of the macro is $\sigma_x\gtrapprox6\times10^{-9}cm^2$, then under atmospheric conditions conducive to lightning (eg. a thunderstorm) the plasma channel would be sufficient to seed a lightning strike with a single leader. This is entirely unlike ordinary bolt lightning in which a long sequence of hundreds or thousands of few-meter-long leaders are strung together. This macro-induced lightning would be extremely straight, and thus highly distinctive. Neither wind shear nor magnetohydrodynamic instabilities would markedly spoil its straightness. The only photographically documented case of a straight lightning bolt is probably not straight enough to have been macro-induced.

    We estimate the region of macro parameter space that could be probed by a search for straight lightning from the number of thunderstorms happening on Earth at any time. We also estimate the parameter space that can be probed by carefully monitoring Jupiter, e.g. using a Jupiter probe.

    All code and data is available at \url{https://github.com/cwru-pat/macro_lightning}.

\end{abstract}

\maketitle


\section{Introduction} 
\label{sec:introduction}

    Assuming General Relativity is the correct theory of gravity on all scales, there is considerable evidence for dark matter \citep{Tanabashi2018}. Despite detailed searches, dark matter has yet to be found at small cross-sections and so we consider larger cross-sections and also larger masses.
   Contrary to widespread misconception, dark matter need not have a small cross-section. The important quantity is $\sigma/m$ -- the ratio of the interaction cross-section of the dark matter (with itself, with baryons, with photons, ...) to the mass of the dark matter candidate.  WIMP dark-matter candidates would achieve a low $\sigma/m$ with a small $\sigma$; primordial black holes would achieve it with a large $m$.
    Macroscopic dark matter (macros) is a broad class of dark-matter candidates that represents an alternative to conventional particle dark matter with wide ranges of masses $M_x$ and large cross sections $\sigma_x$ that could still provide all of the dark matter \citep{jacobs2015macro}.

    Macros typically refer to a family of composite dark matter models arising from some early-universe phase transition, often composed of strange quark matter. Of particular interest would be macros of approximately nuclear density satisfying the geometric cross section (\ref{app:cross_sections})
    \begin{equation}\label{eq:macro_density}
        \sigma_x \approx 2\times 10^{-10} \left(\frac{M_x}{g}\right)^{\frac{2}{3}} [\rm{cm}^2]\,,
    \end{equation}
    as several models for macros describe potential candidates with approximately that density \citep{Sidhu2020reconsider}. The idea that macros could be formed entirely within the Standard Model was originally proposed by \citet{Witten1984} in the context of a first-order QCD phase transition. Though subsequent calculations show this specific mechanism to be unlikely, many other possible mechanisms have been identified. For instance, \citep{Lynn1990, Lynn2010} described a more realistic model for Standard-Model macros as bound states of nucleons with significant strangeness. Nelson \citep{Nelson1990iu} studied the formation of nuggets of strange-baryon matter during a second QCD phase transition -- from a kaon-condensate phase to the ordinary phase. Others have considered non-Standard-Model versions of such objects and their formation \citep{Zhitnitsky2003}.
    In all such models, the interaction cross-section of a macro with another macro, and with baryons, electrons, or photons, is approximately equal to the geometric cross-section.  We adopt this equality as our working hypothesis.

    Some of us, working with colleagues, have recently explored which regions of macro parameter space remain unprobed \citep{jacobs2015macro, jacobs2015resonant, Sidhu2019death, Sidhu2019bolide, Sidhu2020reconsider}. A longstanding constraint comes from examination of a slab of ancient mica for tracks that would have been left by the passage of a macro moving at the typical speed of dark matter in the Galaxy. This was used to rule out macros of $M_x \leq 55\,$g for a wide range of geometric cross sections (see \citet{Price1988ge, DeRujula1984axn, jacobs2015macro}). Various microlensing experiments have constrained the dark-matter fraction  for masses $M_x \geq 10^{23}\,$g \citep{Alcock2001, Griest2013, Tisserand2007, Carr2010, Niikura2019}. \citeauthor{Wilkinson2014angular} \citep{Wilkinson2014angular} utilized the full Boltzmann formalism to obtain constraints from macro-photon elastic scattering using the first year release of Planck data. More recently, the existence of massive white dwarfs was used to constrain a significant region of macro parameter space \citep{Graham2018} (as revisited and extended by \citet{Sidhu2020reconsider}). The region of parameter space for which macros produce injuries similar to a gunshot wound was recently constrained by historical analysis of a well-monitored segment of the population \citep{Sidhu2019death}.

    The parameter space for electrically charged macros, with the macro charge as an additional free parameter, was recently constrained \citep{Sidhu2020charge} based on a variety of terrestrial, astrophysical and cosmological measurements. The parameter space for antimatter macros was constrained by \citet{Sidhu2020anti} using arguments analogous to those cited above.

    More work has been done recently to identify additional ways to probe macro parameter space. With colleagues, some of us have proposed \citep{Sidhu2018auv} using current Fluorescence Detectors that are designed to study High Energy Cosmic Rays, such as those of the Pierre Auger Observatory \citep{Abraham2010}. Separately, we have suggested \citep{Sidhu2019granite} that, for appropriate $M_x$ and $\sigma_x$, the passage of a macro through granite would form long tracks of melted and re-solidified rock that would be distinguishable from the surrounding granite. A citizen-science search for such tracks in commercially available granite slabs is planned to begin through the Zooniverse website sometime later this year. We have also identified the region of parameter space excluded by the null observation of fast-moving meteors ("bolides"), which should have been produced by sufficiently large and fast-moving macros and observed by either of two bolide-observing networks \citep{Sidhu2019bolide}. We determined the region of parameter space that will be probed by planned expansion of the network that is still operating.

    In these works concerning non-anti-baryonic neutral macros, energy is considered to be deposited in matter by the passing macro primarily through elastic scattering. Unlike particle-dark-matter candidates, macros generically interact strongly with matter, so the elastic scattering cross-section is approximately the geometric cross-section. In this case, the energy deposited by a macro transiting the atmosphere 
    \begin{equation}\label{eq:dedx}
        \frac{dE}{dx} = \sigma_x \rho v_x^2\,,
    \end{equation}
    where $\rho \sim 1\,$ kg m$^{-3}$ is the density of the atmosphere at ground level, $\sigma_x$ is the geometric cross-section of the macro, while $v_x$ is its speed. As $v_x$ is much greater than the thermal velocity of air molecules, typical momentum transfers will be of order $m_b v_{x}$ and energy transfers of order $m_b v_{x}^2$.

    The speed of a macro traveling through the atmosphere is thus expected to evolve as
    \begin{equation}\label{eq:atmo_velocity}
        v(x) = v_{0} e^{-\langle \rho \Delta\rangle \sigma_x/{M_x}}\,,
    \end{equation}
    where $\langle \rho \Delta\rangle$ is the integrated column density traversed along the macro trajectory from the point of entry to the location $x$. This will determine the maximum reduced cross-section $\sigma_x/M_x$ expected to deposit sufficient energy to produce an observable signal without being slowed excessively. In previous works e.g. \citet{Sidhu2019death, Sidhu2019bolide}, this limiting value for macros that are interacting at the bottom of the atmosphere was found to be $\frac{\sigma_x}{M_x} \sim 10^{-4}\,$cm$^2$g$^{-1}\,$. This will serve as an upper bound for all Earth-based projections derived in this manuscript.

    One may expect that macros with sufficiently large reduced cross-section would be captured by a planet or star and alter the density distribution of the planet or star. However, macros of interest, i.e. macros with parameters that are still allowed to provide all the dark matter, would be far denser than atomic density and sink to the center of the star or planet. Given the size of these macros, the total density perturbation may have escaped notice.

    As in previous work, we consider macros of a single mass and cross-section, even though a broad mass distribution is a reasonable possibility in the context of a composite dark-matter candidate.

    In this manuscript, we consider the possibility that a macro transiting the atmosphere during the appropriate atmospheric conditions (e.g. a thunderstorm) would initiate an unusual, extremely straight lightning strike. We identify the range of macro parameter space over which that is likely, and consider the possibility that the one documented observation of an abnormally straight lightning strike was triggered by the passage of a macro.  We determine the range of parameter space that could be probed by monitoring the Earth, as well as by observing the atmospheres of Jovian planets, which could probe higher macro masses than any terrestrial detector. 

    The rest of this paper is organized as follows. In Section II we present a review of our current understanding of lightning initiation. In Section III, we discuss the formation of a plasma trail by a passing macro. In Section IV, we calculate the rates of a macro-induced signal. In Section V, we discuss the formation of straight lightning induced by the passage of a macro through the atmosphere. In Section VI, we discuss the observation of a bright UV signal produced by the passage of a macro through a Jovian planet atmosphere. We conclude, with some discussion in Section VII.



\section{A Lightning Review} 
\label{sec:a_lightning_review}

    While the detailed physics of lightning remains a matter of investigation, the broad strokes are well understood. Lightning is an electrical discharge between two regions of large potential difference. Lightning strikes can be classified by the start-end point pair, and sub-classified by the order and charges of those points. For instance, the main classes of lightning are intra-cloud, inter-cloud, cloud-air, and cloud-ground. All except cloud-air lightning may occur in reverse order, like ground-to-cloud or cloud-to-ground. We restrict ourselves to cloud-ground strikes, which are the easiest to observe. The description that follows is almost entirely drawn from the excellent review by Dwyer and Uman \citep{DwyerUman2014}, which should be consulted if a more detailed review of the basic physics is desired.

    A lightning strike is actually two events: first, an ion channel is created from point A to point B, and second, energy flows from B to A. The latter is what is actually observed as ``lightning," and is the luminous signal of the former. The creation of the ion channel under ordinary conditions is a discrete stochastic process of the formation of ``stepped leaders," where a cylindrical atmospheric volume -- ``step" -- is ionized. Each of these steps creates one straight segment of the total jagged lightning bolt. Each step is estimated to take at most $1 \; \mu\rm{s}$. These steps are short compared to the cloud-ground distance -- cloud-level steps are just $\sim10 \rm{m}$, while ground-level steps near $50 m$. The interstep interval ranges from $\sim50 \mu\rm{s}$ at cloud-level to $\sim10 \mu\rm{s}$ at ground. Crucially, the leader persists long after it takes its next ``step". In other words, the path of an organic lightning strike is formed in a series of discrete steps, creating a long but jagged ionized trail that dictates the shape of the resulting lightning strike.

    The propagation direction and charge type of the leader determines the lightning sub-class. For cloud-ground strikes there are four varieties: downward / upward - negative / positive. Thunder clouds are typically negatively charged at the bottom and positively charged on top. Flat ground has regions of differently signed net charge. In a downward-positive strike a positive leader starts near cloud top and steps down to a negatively charged region of ground, a few km below. For all cloud-ground strikes the full channel creation process takes approximately $20$ ms.

    The typical stepped leader has 5 Coulombs of free electrical charge, or $\sim10^{-3} \, \rm{C/m}$. While the leader has a luminous diameter between 1 and 10 m, it is thought to have a conducting core of plasma a few centimeters in diameter. This core acts as a conducting channel, and it is through it that much of the energy flows. Therefore, if a similar quantity of atmospheric charge were to be liberated by other means along a channel of similar dimensions, the resulting ion trail could allow current to flow. This could serve as the basis for a new lightning strike, assuming the trail was created in a region with a sufficient potential difference between connected regions.

    \subsection*{Why Lightning is Jagged} 
    \label{sub:why_lightning_is_jagged}

        Assuming each step in the stepped-leader process has a random azimuthal angle, the probability that for every one of N steps the direction is within $\theta$ degrees displacement from the plane defined by the observer and first step is $\left(\frac{\theta}{180}\right)^{N}$. For illustration purposes only, consider a series of 10 steps  -- and macro-induced lightning should be much longer than that -- in which the maximum step-to-step deviation from a straight path is 10 degrees -- easily observed and much larger than what is expected for macros.
        The probability that this 10-step section of lightning is ``straight'' purely by chance is $3 \times 10^{-13}$. Again, we have underestimated the number of steps, N, and overestimated the maximum allowed degrees of displacement. This accords with observations that straight lightning is very rare and requires no special techniques to detect.
        



\section{Macro-induced \textbf{}Lightning} 
\label{sec:macro_induced_lightning}

    An astrophysical phenomenon somewhat analogous in this context to macroscopic dark matter is cosmic rays. Numerous meteorologists have proposed cosmic rays as a lightning initiation mechanism \citep{Babich2012}. Many contentions, for instance those of Prof Dwyer \citep{scientific_american_2008}, are based on the seeming incompatibility between the small cross section of cosmic rays and the comparatively large ionization channels seen in lightning. Macroscopic dark matter obviates this concern by naturally having a large cross section.

    In most artificially triggered lightning experiments, such as those at the International Center for Lightning Research and Testing (ICLRT) \citep{Hill2012, Hill2013}, a rocket trailing a grounded triggering wire is launched when the quasi-static electric field at ground exceeds $E_{threshold} = 5\,$kV m$^{-1}$ and the flash rate becomes relatively low. In about half of all such launches, an initial stage is successfully triggered, consisting of a sustained upward positive leader typically several kilometers in length followed by an initial continuous current. Often, the initial stage is followed by one or more leader/return stroke sequences, similar to subsequent strokes in natural lightning \citep{Wang1999, rocket2012}.

    The formation of a lightning strike caused by the passage of a macro through the atmosphere is dependent on the formation of a plasma trail produced by the macro scattering elastically off the atoms and molecules. This trail would ``lock in" the lightning-leaders, which serve as the channel through which the charge is transferred in a lightning strike. The plasma trails produced by the macro are similar to the trailing grounded wires as both are sources of free electrons.

    We describe in this section the conditions under which a macro produces a sufficiently large and long-lived plasma channel. We then identify the the ways in which macro-induced lightning differs from natural lightning, in particular in being extremely straight, and so can be used as a signature to search for macros. Finally we discuss the one photographically documented straight lightning bolt.

    \subsection{Forming Plasma Channels} 
    \label{sub:macro_induced_plasma_channels}

        We review the key quantities about this plasma first; we refer the reader to reference \citep{Sidhu2018auv} for more details. Due to the longevity of lightning leaders, we need only demonstrate that the macro channels contain as much charge density as a natural leader, and persist long enough for the lightning leader to ``lock in" along the macro path.

        \citeauthor{Cyncynates2016} \citep{Cyncynates2016} considered the formation of plasma channels by macros passing through rock. For passage through the atmosphere, additional cooling terms come into play: radiative cooling, expansion cooling, and turbulent mixing \citep{Picone1983}. Ignoring these for the moment, we can propagate the initial energy deposition by the macro outward radially away from that trajectory using the heat equation.
        The temperature field after some time $t$ is
        \begin{equation}\label{eq:temperature_field}
        	T(r,t) = \frac{\sigma_{x} v_x^2}{4\pi \alpha c_p}\frac{e^{-\frac{r^2}{4t\alpha}}}{t},
        \end{equation}
        where $\alpha \approx 10^{-4}\,$m$^2\,$s$^{-1}$exp$(D/10$km$)$ is the thermal diffusivity of the air, and $c_p \approx 25$ kJ kg$^{-1}$ K$^{-1}$ is the specific heat of the air \citep{Capitelli2000} (The specific heat varies around a mean of $\sim25\,$kJ kg$^{-1}\,$K$^{-1}$ for temperatures between $10^4\,$K and $10^5\,$K).

        We invert \eqref{eq:temperature_field} to obtain $\pi r_I(t)^2$, the area at time $t$ that has reached a particular state of ionization $I$ characterized by the appropriate ionization temperature $T_I$. We do this by setting $T(r,t) = T_I \approx 5\times10^4$K \citep{EisazadehFar2011}, sufficient to ionize the 2p electrons of N and O. This area is given by
        \begin{equation}\label{eq:ionization_crosssection}
            \pi r_I(t)^2 = 4\pi\alpha t\log\left(\frac{\sigma_{x} v_x^2}{4\pi \alpha t c_p T_I}\right) .
        \end{equation}
        According to \eqref{eq:ionization_crosssection}, after the macro passes, the size of the ionized region grows to a maximum of
        \begin{equation}\label{eq:maximum_area}
            A^{\rm max} \equiv \pi (r_{I}^{\rm max})^2=\frac{\sigma_x v_x^2}{e  c_p T_I} \approx 7.5\!\times\!10^{3}\sigma_x \,\left(\frac{v_x}{250\frac{\rm{km}}{\rm{s}}}\right)^2\,.
        \end{equation} 
        This happens at 
          \begin{equation}\label{eq:cooling_time}
            t_{I}^{\rm max}=\frac{\sigma_x v_x^2}{4\pi e \alpha c_p T_I} \approx 6\,s\left(\frac{\sigma_x}{cm^2}\right)\left(\frac{v_x}{250kms^{-1}}\right)^2e^{-\frac{D}{10 km}}\,.
        \end{equation} 
        It then shrinks back to $0$ at $t_I^0=e~t_{I}^{\rm max}$.

        It is important to note that $A^{\rm max}$ is independent of $\alpha$. Turbulent mixing and expansion cooling will change the value of $\alpha$, and change the precise temperature profile \eqref{eq:temperature_field}, but will not change the maximum number of ionized atoms per unit length along the macro trajectory. They could in principal cool the plasma too quickly to allow leader formation along the channel. This will be considered in the following subsections.

        Radiative cooling could have more deleterious effects by removing the energy in the plasma to a distant location, too far to participate in the leader initiation. However, in \citet{Sidhu2018auv} we show that the effects of radiative cooling are negligible for $\sigma \lesssim 3\times 10^{-3}{\rm cm}^2$. For larger $\sigma$, the linear charge density in the macro channel saturates.


    \subsection{Inducing Lightning Leaders} 
    \label{sub:inducing_lightning} 

        In order to initiate lightning, we need to create charged filaments with linear charged densities sufficient to seed a leader. In natural lightning, the leaders have \citep[][p. 152]{DwyerUman2014} a linear electron density $\lambda_e^{\rm natural}\simeq 6\times 10^{13} cm^{-1}$. By comparison, within the plasma channel at time $t_I^{\rm max}$ the linear free-electron density will be
        \begin{equation}
            \lambda_e^{\rm macro} \simeq \pi (r_I^{\rm max})^2 n_a f_e
        \end{equation}
        where $n_a$ is the number density of atoms in air, and $f_e$ is their ionization level. Taking $f_e\simeq0.5$ appropriately accounts for the fact that the 2p electrons of N and O are ionized at $T_I$ but the 1s and 2s electrons are not.

        Knowing that each luminous step leader propagates \citep{DwyerUman2014} in at most $1\mu{s}$, followed by a pause of between $50\mu{s}$ (at high altitude) and $10\mu{s}$ (near the ground) between leaders, we therefore require that
        \begin{equation}\label{eq:tI0min}
            t_{I}^0 \geq 1\mu{s} \quad \implies \quad \sigma_x > 6\times 10^{-8}cm^2\,,
        \end{equation}
        and that the linear charge density in the macro-induced plasma trail
        \begin{equation}\label{eq:lambdaemin}
            \lambda_e^{\rm macro}\geq \lambda_e^{\rm natural}
            \quad \implies \quad \sigma_x > 10^{-8}cm^2 \,.
        \end{equation}
        
        Equation \eqref{eq:tI0min} is more stringent than \eqref{eq:lambdaemin}; however, $1\mu{s}$ is an upper bound for the time-scale over which each step leader forms, and represents propagation along the step leader at approximately $0.05c$. Positive return strokes travel \citep{Idone1987} at $c/3$, which may be a more realistic estimate of the propagation speed. This would drop the minimum applicable $\sigma_x$ to $10^{-8}cm^2$. Nevertheless we quote our accessible macro parameter space using the more restrictive $\sigma_x \geq 6\times 10^{-8}cm^2$.
        
        Equations \eqref{eq:tI0min} was calculated using the diffusive cooling of \eqref{eq:temperature_field}, but turbulent mixing is known to be a more effective cooling mechanism. Likewise, the expansion cooling will strongly cool the macro channel, impacting $\alpha$ in \eqref{eq:cooling_time}. In order to induce a leader, the plasma channel needs to exist long enough for a leader to connect to the macro's plasma channel. Considering the average speed of a lightning leader, this might appear to be a problem. In ordinary lightning the \textit{inter}-step pauses are long; these dominate the propagation time and lower the average propagation speed of the leader. However, this is the wrong timescale to consider. Leaders require these long inter-step intervals because there is not already a highly-ionized channel to follow. As reference, \citeauthor{Betts2003} \citep{Betts2003} uses a 0.2mm copper wire to induce leader formation. Our values of $\sigma$ yield macro channels with at least 5000 times greater linear charge density (see \ref{sub:staying_straight}). Given this highly-ionized channel, the leaders can propagate at the \textit{intra}-step velocity, which is approximately $c/3$, for which the relevant time step is $1 \mu s$.
        Turbulent mixing acts only on timescales of $300\mu{s}$ \citep{Picone1983}, which is much longer than the $1\mu{s}$ required. Other cooling mechanisms are similarly unimportant.


    \subsection{On Cloud Structure and Charge Exchange}

        Clouds have complicated charge structure. Broadly speaking, clouds are net neutral with a main positive charge (often on top), and a main negative charge. The regions are roughly 10 km tall, for a total cloud complex of order 20 km. The ground charge, up to 4 km distant, is effectively an induced charge from the cloud. Smaller charge regions are common, for instance beneath the main bottom charge, with a corresponding dipole on the ground.
        
        Lightning occurs between effectively all regions, including the ``small" charge ones. Lighting acts to neutralize the local electric field, negating the potential difference between the two regions. When a macro passes through a cloud, as it transits each charge region it will induce intra-cloud lightning. We note that this lightning will have all the same signatures as groundstrokes, discussed in \ref{sec:signatures_of_macro_induced_lightning}, but is not the focus of this paper since the obscuration by clouds might make analysis challenging. Intra-cloud lightning \textit{might} -- stressing that this is speculative -- prevent groundstrokes if the macro passes through a small charge region between the main bottom charge and ground. In this scenario, the small charge region might be neutralized and the charge in cloud and ground is now of the same sign. Not enough is known to determine whether the charge neutralization is sufficient to prevent groundstrokes, or if this picture is even correct.
        
        The frequent case that the macro passes from the main bottom charge straight to ground is much simpler. When the macro pierces cloud bottom, the induced intracloud lightning occurred roughly 10 km and 40 ms away. At ground connection that distance can be 14 km and 60 ms distant. Given that small charge regions regularly produce lightning, the main charge region will not be depleted by the earlier and distant discharge.

        We now have two scenarios: groundstroke from main bottom charge to ground and groundstroke from small charge region to ground. The former should produce groundstrokes, the latter \textit{might} not. To approximate the uncertainty in the occurrence rate of the latter case and how efficiently it prevents groundstrokes we divide our lightning generation rate by an extremely conservative factor of 2.
        
        There is a third important scenario to consider. Instead of traveling from cloud to ground, a macro can pass through a cloud upward after traveling through the Earth. This causes ground-cloud lightning. We calculate later (\ref{eq:twicelightning}) the relevant cross section for a macro to be able to pass through the Earth and still have speed sufficient to cause distinctive lightning (\ref{sec:signatures_of_macro_induced_lightning}).


    \subsection{Signatures of Macro-induced Lightning} 
    \label{sec:signatures_of_macro_induced_lightning}

        Our macro-induced lightning initiation model differs from \citet{DwyerUman2014} in a few important regards. Let us review the differences thus far. First, as the macro trail acts as a ``pre-leader", the leader-creation process is not stochastic but deterministic, with normal lightning leaders ``locking in" along the macro channel. Second, since the macro constantly creates the plasma channel the leader propagates continuously along this channel. The mode of the macro velocity distribution, 250 km/s, is near exactly the propagation velocity of the leaders ($200\,$ km s$^{-1}$), when including the interstep interval. However, the propagation of the leader within each step is known to take at most $1\mu$s, and therefore to be at a velocity of at least $10^4\,$ km s$^{-1}$, and may perhaps be as much as the $c/3$ measured for positive return strokes. So as the macro continuously creates a plasma trail the leader will propagate at this same velocity. Thus in macro-induced lightning leaders are continuous, not discrete.

        \begin{figure}[ht]
            \centering
            \includegraphics[width=\linewidth]{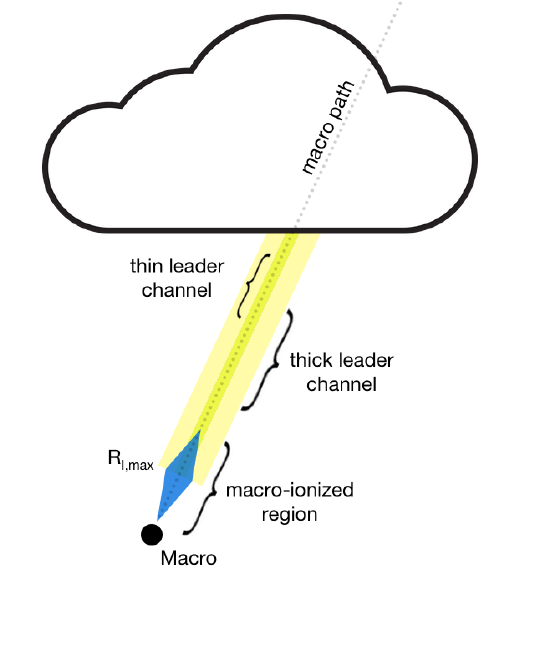}
            \caption{\textit{(Not to scale)} Graphic representation of macro plasma channel seeding continuous leader. Channel direction can also be from ground to cloud. Macro plasma trail expands to maximum area $A^{\rm{max}}$ before cooling. As $v_{\rm{step}} < v_{\rm{macro}} < v_{\rm{leader}}$, the lightning-leader takes no ``steps", instead propagating continuously with the macro trail.}
            \label{fig:macro_graphic}
        \end{figure}

        This offers a few testable predictions: the leader process produces no light pulses during steps, the RF and X-ray signatures of the leader steps are similarly different. However these predictions are harder to test since different types of lighting are observed to have distinct RF and X-ray signatures \citep{Hare2020}. The most conspicuous prediction is that {macros source abnormally straight lightning} compared to the typical lightning strikes observed.

        We note some \textbf{caveats}. First, the stepped leader model does not apply for the last tens of meters as the ground emits an upward propagating stepped leader which will connect to the downward propagating plasma channel. Moreover, for macros moving slower than $250\,$km/s, the lightning is expected to be jagged like regular lightning as the stepped leader would eventually overtake the macro trail. For macros moving significantly faster than $250\,$km/s the lightning is expected to be straight the entire pathway from cloud to ground as the ground will not have time to emit or significantly propagate its own stepped leader. 

        Since macros are expected to move according to a Maxwellian velocity distribution in a frame co-moving with the Galaxy,
        \begin{equation}\label{eq:maxwellian}
        	f_{MB}(v_x) = 
        		\frac{4\pi v_x^2}
        		{\left({\pi v_{vir}^2}\right)^{3/2}}~
        		e^{-\left(\frac{v_x}{v_{vir}}\right)^2}, 
        \end{equation}
        where $v_{vir} \approx 250~ \text{km s}^{-1}$ \footnote{
        	This is the distribution of macro velocities in a non-orbiting frame moving with the Galaxy. When considering the velocity of macros impacting the atmosphere, \eqref{eq:maxwellian} is modified by the motion of the Sun and Earth in that frame, and by the Sun's and Earth's gravitational potential. We have taken into account these effects (as explained in \citet{Freese2013}), except the negligible effect of Earth's gravitational potential.

        	Recent hydrodynamical simulations of Milky Way-like galaxies including baryons, which have a non-negligible effect on the dark matter distribution in the Solar neighborhood \citep{Tanabashi2018} have been performed to determine the correctness of assuming a Maxwellian distribution. These simulations find that the velocity distributions are indeed close to Maxwellian. As discussed previously, macros are expected to move according to \eqref{eq:maxwellian}. Taking this minimum speed requirement into account, we find that $71\%$ of all macros in the distribution will be moving at at least $250\,$km/s. 
    	}.
        and taking the relative motion between the macro and Earth into account, we find that $71\%$ of all macros in the distribution will be moving at at least 250 km/s.

        Additionally, we expect that the mechanism outlined here may not hold true if the macro comes in at a trajectory that is mostly parallel to the ground. There is a critical angle at which a macro trail is sufficiently misaligned from the storm electric field such that the electric field induces offshoot lightning channels, obviating the straight-lightning prediction. This is poorly constrained because plasma channels in air are analogous to wires surrounded by an insulator. The breakdown voltage is highly dependent on atmospheric properties such as moisture and particulate content, etc. Despite this, order of magnitude calculations suggest the critical angle is approximately unity. As example, considering a cloud-to-ground macro-induced plasma channel for a critical angle of 30$\degree$ from a perfectly perpendicular trajectory, $25\%$ of all macro trajectories would fall in this cone. We use this number this when calculating the maximum mass that could be probed by a careful monitoring of thunderstorms on Earth.



    \subsection{Staying Straight} 
    \label{sub:staying_straight}

        Although a macro creates a straight plasma channel, at least two mechanisms will spoil that: the $m=1$ MHD instability on small scales and wind shear on large scales. Of these only the wind-induced non-linearity is expected to be observable by commercial-grade equipment. We discuss both.

        There have been a number of studies investigating how to artificially induce lightning strikes through laser-generated plasma channels \citep[see][]{Kasparian2008}. Though no strikes have yet been directly triggered due to technical limitations in producing a continuous ground-to-cloud channel. Instead, an informative analogue to macro-induced lightning is lightning induced by charged particles from the IVY-MIKE 1952 nuclear explosion test on Enewetak Atoll \citep{IVY-MIKE1987}.

        In laboratory tests to simulate the IVY-MIKE lightning, laser-guided electric discharges were used to create a ${\sim 1}$ m straight plasma filament, radius $R_f\lesssim 1$ cm, within a reduced density channel, radius $R_d \lesssim 2$ cm \citep[][fig. 6]{IVY-MIKE1987}. On timescales exceeding $40 \, \mu s$, the $m=1$ magnetohydrodynamic (MHD) mode kinks the central filament, with perturbations of amplitude $R_e$ ($R_f < R_e < R_d$) and growing wavelength $\lambda$. $R_f$,\, $R_e$ and $R_d$ grow sub-linearly \citep[][fig. 9]{IVY-MIKE1987}. Extrapolating to 20 ms (ground-to-cloud time), the radius of the reduced density channel is $R_d<3m$. After 1 ms, the central filament kink radius $R_e$ has nearly plateaued at 10 cm, while the filament $R_f$ itself is stable at 1-2 cm. The $m=1$ mode wavelength is $\lambda\simeq4 m$. These lab-measurements of $R_d$ are consistent with observed lightning. While the amplitude and wavelength of the kink mode explain why it has yet to be observed. The $m=1$ instability should not alter the apparently straight lightning path, which is observed as the reduced density channel.

        Wind shear is not expected to introduce significant long wavelength deviations from straightness. The typical timescale of cloud-to-ground ion channel formation is ${\approx 20}$ ms. The return stroke, aka the first lightning strike \citep{DwyerUman2014} occurs directly following the ion channel creation and propagates at $c/3$ \citep{Idone1987}. At a wind speed of ${\approx 20}$ m/s \citep{Choi2004}, high for the typical thunderstorm, local regions of the plasma channel can be transported by ${\sim 0.5}$ m. Even if wind shear transports neighboring plasma channel components in opposite directions, the observed deviation from a straight strike is just $1$m. Repeated strikes are generally separated by ${\sim 50}$ ms, contributing a further ${\sim 2}$ m deviation of the channel. In actuality, repeated strokes can be distinguished by any camera with $>30$ fps. These effects should not contribute significantly on the first strike and a macro-induced lightning track is predicted to be nearly perfectly straight.
        
        There is one further effect to consider: whether the leader will break out from the macro channel. For instance, in triggered lightning experiments where a conducting wire is sent up to charge clouds, the discharge largely follows the wire, but consistently breaks away at some point. Typically, these conducting wires are made of copper, which has a conduction-electron density of $\simeq 4 \times 10^{13} cm^{-3}$. We have required the macro channels to have a linear free-electron density $\lambda_e^{macro} \geq \lambda_e^{natural} \simeq 6 \times 10^{13} cm^{-1}$. Those channels have a maximum radius of 1-2 cm. Therefore the channels have conduction-electron densities greater than or comparable to copper. More importantly the linear electron conduction-density is \textit{much} greater than the wires that are used in induced lightning experiments. For example, \citeauthor{Betts2003} \citep{Betts2003} used 0.2mm copper wire, with $\lambda\simeq 10^{10} cm^{-1}$. This is at least 5000 times less than a macro channel. Breakout is much less likely.

        Considering the $m=1$ instability, wind shear, and breakout probability, deviations from straightness by even 1 reduced density channel width at any point along the channel path disqualifies a lightning event as a macro-induced event candidate. Considering the probabilities associated with naturally straight lightning -- section \ref{sub:why_lightning_is_jagged} -- any extremely straight lightning seems an excellent candidate for further analysis.




\section{Macro Search and Potential Constraints}  
\label{sec:macro_search_and_constraints}

    Using the distribution \eqref{eq:maxwellian}, transformed to the solar frame \citep{Freese2013}, the macro flux on a planet would be given by,
    \begin{equation}\label{eq:planet_macro_flux}
        F_{x} = \frac{\rho_{x,0}}{M_{x}} \int v_{x} f_{MB,SS} dv_x,
    \end{equation}
    where $\rho_{x,0} = 5 \times 10^{-25}$ g cm$^{-3}$ is the local DM density \citep{Bovy2012}, $M_{x}$ is the mass of the macro and the integral accounts for the velocity distribution of all macros, and $f_{MB,SS}$ is the Maxwell Boltzmann distribution in the Solar System frame. With this, we calculate the estimated rate of macro-induced lightning strikes
    \begin{equation}\label{eq:macro_lightning_rate}
        n_{ml} = \frac{\rho_{x,0} \pi R_{O}^2 f_{TS} f_{LE}}{M_{x}}\int v_{x} f_{MB,SS} dv_x\,,
    \end{equation}
    where $R_{O}$ is the planet's radius, $f_{TS}$ is the fraction of planet's surface currently experiencing a thunderstorm, and $f_{LE}$ is the fraction of macro strikes in thunderstorms that actually lead to an observable event. For the range of cross-sections of interest, $f_{LE}\simeq1$.

    We note that sufficiently fast-moving and small macros would not be slowed down appreciably even considering the column density encountered passing through the diameter of the Earth. More concretely, using the PREM density profile \citep{Dziewonski1981}, we determined the critical reduced cross-section below which macros traveling initially at above $v\sim 500\,$km s$^{-1}$, which represent half of all the macros in the velocity distribution, would not be slowed down to half their initial velocity on traveling through the Earth. This critical reduced cross-section is found to be
    \begin{equation}\label{eq:twicelightning}
        \sigma_x/M_x \leq 1\times 10^{-10} [\rm{cm}^2\rm{g}^{-1}] \,.
    \end{equation}
    Since the critical angle relative to normal for co-alignment of the macro and storm electric field is probably not oblique (\ref{sec:signatures_of_macro_induced_lightning}), all macros of relevant cross section will have two opportunities to induce lightning.
    In Figure 2, this region of the parameter space that could be probed is shaded in gray (and labeled "upward") to show this. The reason that this behavior is important is that such macros will have two opportunities to initiate straight lightning just before entering the Earth and just after exiting. Even more significant, is that macros satisfying this expression would be able to initiate lightning from ground-to-cloud, which would be even more compelling observation for the existence of such objects.

    \subsection{Straightest Observed Lightning} 
    \label{sub:straightest_observed_lightning}

        We conducted a search in the physics literature and publicly available new sources for reports of anomalously straight lightning. The most promising candidate was reported in Mutare, Zimbabwe on 15 February 2015 \citep{Zimbabwe} and recorded at 30 frames per second with a Panasonic Lumix DMC-TZ10 compact camera in scene mode. The observed lightning strike is a cloud-ground strike with no secondary strikes. The maximum projected deviations from perfect linearity are of order a few diameters. As the thickness of a beam of lightning is between 1m and 10m (and does not depend significantly on the considered macro parameter space), even this straight lightning strike is mostly likely not straight enough to have been induced by a macro.

        The expected signature from a macro-induced lightning strike would be very unique. This presents, in theory, a straightforward way to search for macros by looking for macro-induced lightning strikes, and to place constraints on macros if no such strikes are observed. 


    \subsection{Potential Macro Constraints} 
    \label{sub:macro_constraints_on_earth}

        To place constraints on macros from the non-observation of any straight lightning strikes, we note that the passage of a macro through the area covered by a thunderstorm is a Poisson process. Thus the probability of $n$ passages over a given exposure time, $\Delta t$, $P(n)$ follows the distribution
        \begin{equation}\label{eq:poisson}
            P(n) = \frac{\left({n_{ml} \Delta t}\right)^n}{n!} e^{-n_{ml} \Delta t}\,.
        \end{equation}
        The continued failure to observe a macro-induced lightning strike would allow us to conclude that $n_{ml}\Delta t<3$ at 95$\%$ confidence level.

        To calculate the expected macro-induced lightning rate on Earth, we take $R_{O} = \bar{R_{\bigoplus}} = 6 \times 10^8$ cm. At any given time Earth experiences approximately 2,000 thunderstorms \citep{NatGeo}, with an average $20$ km in diameter, giving $f_{TS}\simeq 0.3\%$.

        With these assumptions, and should we observe 0 very straight lightning strikes in two years, we could place an upper bound on the mass of a macro up to $M_x \sim 10^6\,$g for $\sigma_x\gtrapprox6\times10^{-8}{\rm cm}^2$. The exact projections are shown in Figure 2. It is of particular significance that this method is sensitive to probing the nuclear density line.
        
\begin{figure*}
    \includegraphics[width=\textwidth]{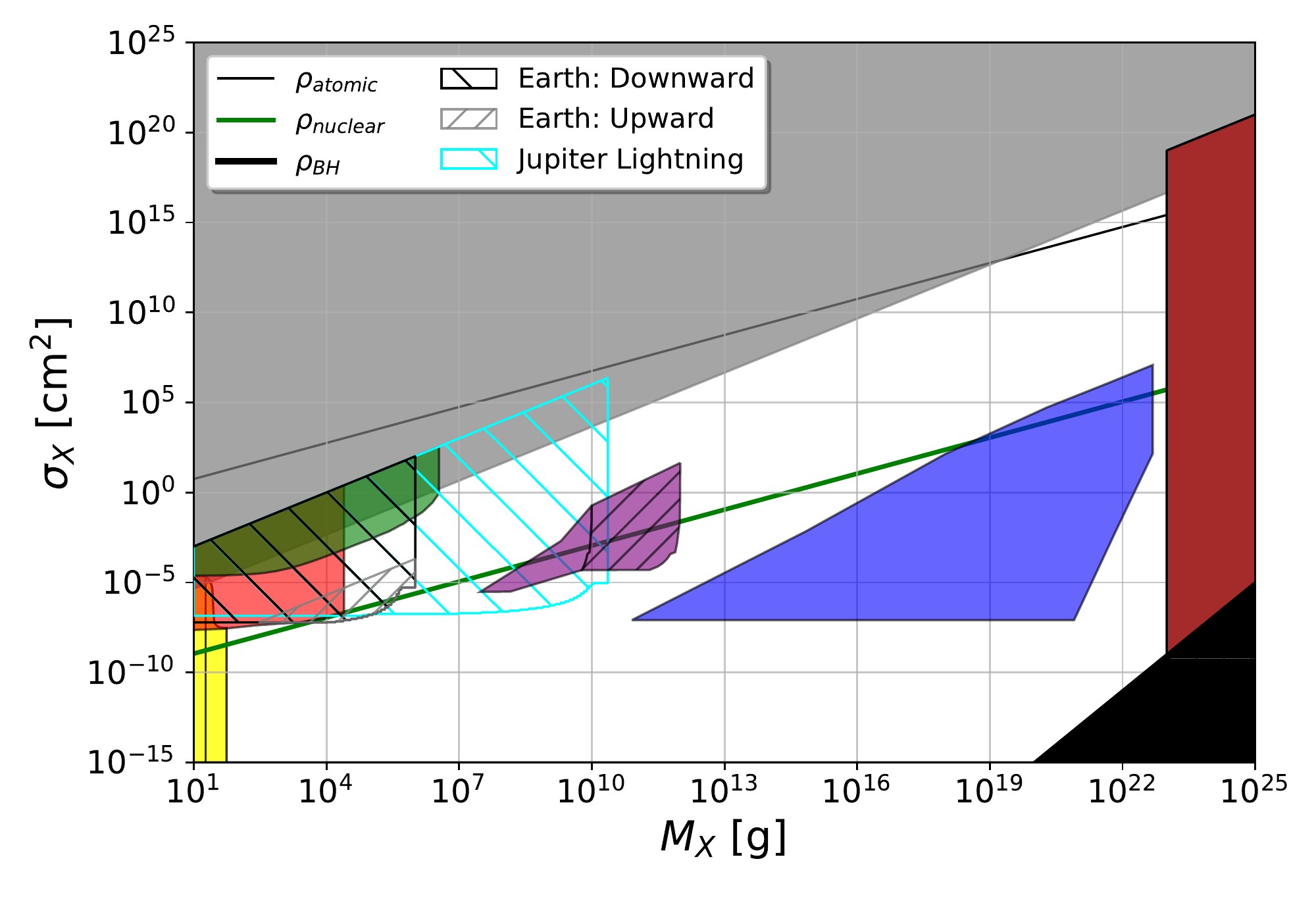}
    \caption{
        Figure 1 of \citep{Sidhu2020reconsider} with the updated potential constraints discussed in the text. Earth lightning projections are in black hatching and Jupiter lightning projections are in cyan hatching. The lower right right constraint curves up because the number density of macros decreases with increased mass which requires a larger fraction of the velocity distribution. The step in the lower right is the average observed minimum macro velocity due to gravitational infall.
        \\
        Objects within the bottom-right corner are excluded as they are denser than black holes of the same mass. The {\color{gray} solid gray} region is ruled out from structure formation \citep{Wilkinson2014angular}; the {\color{yellow} yellow} from mica observation \citep{DeRujula1984axn, Price1988ge};  the light {\color{blue} purple} from superbursts in neutron stars; the {\color{cyan} light blue} from WDs becoming supernovae (\citet{Graham2018} as revised in \citet{Sidhu2020reconsider});  the {\color{red} red} from a lack of human injuries or deaths \citep{Sidhu2019death}; the {\color{green} green} from a lack of fast-moving bolides \citep{Sidhu2019bolide}; the {\color{purple} maroon} from a lack of microlensing events \citep{Niikura2019, Alcock2001, Tisserand2007, Carr2010, Griest2013}. Solid colors denotes verified constraints, hatching for potential constraints.
        The sub-region of Earth constraints shaded in {\color{gray} gray} shows the region of parameter space where macros satisfy \eqref{eq:twicelightning}
    }\label{fig:constraints}
\end{figure*}

        We calculate these potential constraints with the simplification of a gravitational infall velocity determined only by the mass of the Sun and Earth, not accounting for the Earth's orbital velocity. This only noticeably affects the small lower right plateau in the constraint curve of \ref{fig:constraints}, which is determined by this velocity.

        To achieve these constraints requires more detailed observations / reporting of lightning as a significant fraction of lightning is not observed, and only a fraction of those events are recorded. Fortunately, lightning strikes are heavily concentrated over land \citep{Christian2003}, increasing the possibility of establishing a dedicated monitoring program. It also increases the probability of reporting by casual observers, since nearly straight lightning strikes are rare enough to generate press (see \citet{Zimbabwe}).

        For e.g. $M_x = 100\,$g over the range of cross-sections of interest, the macros can make up no more than $2\times 10^{-3}$ of the dark matter \citep{Sidhu2019death}. Thus, we would expect a macro-induced lightning rate of $\sim 10^{-6}$ s$^{-1}$, combining this maximum fraction with the rate \eqref{eq:macro_lightning_rate}. This is already much lower than the actual observed rate of lightning strikes on Earth, which is on order of 50 to $100$ s$^{-1}$ \citep{Mackerras1998}. This implies that we cannot significantly constrain macros as dark matter through lightning rates alone, as the macro-induced lightning signal would always be significantly outnumbered by the rate of regular lightning strikes. However, as discussed in Section \ref{sec:signatures_of_macro_induced_lightning}, the lightning strikes induced by macros are expected to be significantly straighter than regular lightning strikes. Thus, we expect to see straight lightning caused by macros regardless of whether the macros populate a part of parameter space where they can or cannot contribute all the dark matter.

    


\section{Jupiter} 
\label{sec:jovian_bolides}

    Given its size relative to Earth, a search for macros using Jupiter (or another gas giant planet) as the target holds great potential for exploring larger macro masses than can be explored using Earth as the target. Although on Earth we have the advantage of being able to search for the effects of macros on targets like rocks that have ``integrated" for extremely long exposure times \citep{Sidhu2019granite}, that advantage is nullified when looking for transient phenomena such as lightning flashes, that need to be observed in real time. Thus, a potential signal of macros is the production of straight lightning in the Jovian atmosphere as discussed above for Earth's atmosphere. In this section we will briefly discuss some of the potential power and challenges of using Jupiter as a target for macro-induced lightning signals. The major strength of searching for straight Jovian lightning is the size of the target. The surface area of Jupiter is 125 times that of Earth, suggesting that it is a potentially valuable target to search for macro-induced fluorescence or macro-induced lightning. Lightning has been observed near the Jovian poles by every passing satellite. Earlier mysteries as to its origins have recently been clarified based on observations from the Juno mission \citep{Brown2018}, and it is now understood to be be described by essentially the same physics as terrestrial lightning.

    Making concrete claims about the observability of macro-induced lightning on Jupiter is difficult. This is due to two factors: the physics of lightning in Jupiter's atmosphere is even less well-understood than that on Earth, and the logistics of monitoring Jovian lightning is much more difficult given its distance. For example, it is currently unclear why lightning does not form over the entire surface of Jupiter but only the poles. This could reduce the region of parameter space that could be probed through this method. One additional difference is that in the case of Jovian lightning, we are only concerned with cloud-cloud lightning, as opposed to focusing on cloud-ground lightning for Earth. This is because the `ground' for Jupiter is essentially unobservable, and thus it is more useful to look for intercloud lightning strikes in the upper layers of the Jovian atmosphere. In addition, observing the morphology of Jovian lightning presents obvious technical challenges, but could be overcome either by using high-resolution space telescopes from earth, or by using future Jovian weather satellites that will make precise measurements of Jupiter's atmospheric phenomenon. Given that many lightning strikes on Jupiter will be obscured by the cloud cover, it would be more advantageous to use detection methods that do not rely on visual morphology to differentiate macro-induced lightning from organic lightning. For example, as mentioned earlier, one could potentially use the RF signal to differentiate straight lightning bolts without visual confirmation. This could be accomplished using RF instrumentation on existing probes such as JUNO, or proposed upcoming probes such as the Jupiter Ganymede Orbiter \citep{Brown2018}. An exact calculation of sensitivity would require a better estimate of the amplitude of RF signals from macro-induced Jovian lightning, which is outside the scope of this paper.

    Despite these theoretical and observational challenges, we shade, in Figure 2, the region of parameter space that could be probed assuming that lightning occurs only over $10\%$ of the surface of Jupiter, which is likely an underestimate. We also assume lightning physics is identical on Jupiter compared to on Earth, and that this lightning is detectable and distinguishable from non-macro induced lightning using some future technology. We do not claim that our forecasts for constraints due to Jovian lightning are definitive, but instead present them as a potential future area of research, worthy of more in-depth investigation.



\section{Conclusion} 
\label{sec:conclusion}

    In this manuscript, we have proposed that macros could result in the formation of distinctive, abnormally straight lightning that, to our knowledge, has not been documented on Earth. This could serve as the basis for a high-sensitivity search for macros of higher mass and lower geometric cross section than other methods that have been proposed. We also proposed using lightning on Jupiter to probe a much larger region of parameter space, although a detailed consideration of this idea must still be performed.



\medskip
\section{Acknowledgments} 
\label{sec:acknowledgements}

    This work was partially supported by Department of Energy grant de-sc0009946 to the particle astrophysics theory group at CWRU.

    We acknowledge the support of the Natural Sciences and Engineering Research Council of Canada (NSERC) Canadian Graduate Scholarships - Master's Program, [funding reference number 542364 / 2019 (Nathaniel Starkman), 542579 / 2019 (Harrison Winch)]
    
    Nathaniel Starkman received partial support from an Ontario Early Researcher Award (ER16-12-061; PI Bovy)

    \subsection*{Software} 
    \label{sub:software_citation}

        The analyses done for this paper made use of: \texttt{astropy} \citet{code_Astropy2013, code_Astropy2018}, \texttt{IPython} \citet{code_Perez2007}, \texttt{matplotlib} \citet{code_Hunter2007}, \texttt{numpy} \citet{code_Walt2011}, and personal and project-specific packages \citet{code_utilipy2020, code_starkplot2020, code_macrolightning2020}.




\section{Appendix}  
\label{sec:appendix}

    \subsection{Relationship between macro mass, cross-section and internal density}  
    \label{app:cross_sections}
    
        Macroscopic dark matter is much larger than the size of a proton or neutron, and therefore the cross section is both the geometric cross section and the cross section for elastic scattering.
         
        Deriving the cross section with reference to nuclear density,

        \begin{align}
            \rho_X & \propto \frac{M_x}{r_X^3} \quad \rho_{nuclear} \propto \frac{M_{nuclear}}{r_{nuclear}^3}
        \end{align}

        Taking $\rho_{nuclear}=3.6\times10^{14} [\rm{g}\rm{cm}^{-3}]$ and solving for the cross section in terms of the nuclear density.

        \begin{align}
            \sigma_X &= 2.4\times 10^{-10} \frac{\rho_{nuclear}}{\rho_X}^{2/3} \left(\frac{M_x}{M_{nuclear}}\right)^{2/3} \rm{[cm]}^2 \nonumber \\
            &\simeq
            2\times 10^{-10} \left(\frac{M_s}{g}\right)^{2/3} \rm{[cm]}^2.
        \end{align}
    



\bibliographystyle{apsrev4-1}
\bibliography{references}

\providecommand{\noopsort}[1]{}\providecommand{\singleletter}[1]{#1}%
\begin{thebibliography}{59}%
\makeatletter
\providecommand \@ifxundefined [1]{%
 \@ifx{#1\undefined}
}%
\providecommand \@ifnum [1]{%
 \ifnum #1\expandafter \@firstoftwo
 \else \expandafter \@secondoftwo
 \fi
}%
\providecommand \@ifx [1]{%
 \ifx #1\expandafter \@firstoftwo
 \else \expandafter \@secondoftwo
 \fi
}%
\providecommand \natexlab [1]{#1}%
\providecommand \enquote  [1]{``#1''}%
\providecommand \bibnamefont  [1]{#1}%
\providecommand \bibfnamefont [1]{#1}%
\providecommand \citenamefont [1]{#1}%
\providecommand \href@noop [0]{\@secondoftwo}%
\providecommand \href [0]{\begingroup \@sanitize@url \@href}%
\providecommand \@href[1]{\@@startlink{#1}\@@href}%
\providecommand \@@href[1]{\endgroup#1\@@endlink}%
\providecommand \@sanitize@url [0]{\catcode `\\12\catcode `\$12\catcode
  `\&12\catcode `\#12\catcode `\^12\catcode `\_12\catcode `\%12\relax}%
\providecommand \@@startlink[1]{}%
\providecommand \@@endlink[0]{}%
\providecommand \url  [0]{\begingroup\@sanitize@url \@url }%
\providecommand \@url [1]{\endgroup\@href {#1}{\urlprefix }}%
\providecommand \urlprefix  [0]{URL }%
\providecommand \Eprint [0]{\href }%
\providecommand \doibase [0]{http://dx.doi.org/}%
\providecommand \selectlanguage [0]{\@gobble}%
\providecommand \bibinfo  [0]{\@secondoftwo}%
\providecommand \bibfield  [0]{\@secondoftwo}%
\providecommand \translation [1]{[#1]}%
\providecommand \BibitemOpen [0]{}%
\providecommand \bibitemStop [0]{}%
\providecommand \bibitemNoStop [0]{.\EOS\space}%
\providecommand \EOS [0]{\spacefactor3000\relax}%
\providecommand \BibitemShut  [1]{\csname bibitem#1\endcsname}%
\let\auto@bib@innerbib\@empty
\bibitem [{\citenamefont {Tanabashi}\ \emph {et~al.}(2018)\citenamefont
  {Tanabashi} \emph {et~al.}}]{Tanabashi2018}%
  \BibitemOpen
  \bibfield  {author} {\bibinfo {author} {\bibfnamefont {M.}~\bibnamefont
  {Tanabashi}} \emph {et~al.} (\bibinfo {collaboration} {Particle Data
  Group}),\ }\href {\doibase 10.1103/PhysRevD.98.030001} {\bibfield  {journal}
  {\bibinfo  {journal} {Phys. Rev. D}\ }\textbf {\bibinfo {volume} {98}},\
  \bibinfo {pages} {030001} (\bibinfo {year} {2018})}\BibitemShut {NoStop}%
\bibitem [{\citenamefont {Jacobs}\ \emph {et~al.}(2014)\citenamefont {Jacobs},
  \citenamefont {Starkman},\ and\ \citenamefont {Lynn}}]{jacobs2015macro}%
  \BibitemOpen
  \bibfield  {author} {\bibinfo {author} {\bibfnamefont {D.~M.}\ \bibnamefont
  {Jacobs}}, \bibinfo {author} {\bibfnamefont {G.~D.}\ \bibnamefont
  {Starkman}}, \ and\ \bibinfo {author} {\bibfnamefont {B.~W.}\ \bibnamefont
  {Lynn}},\ }\href {\doibase 10.1093/mnras/stv774} {\bibfield  {journal}
  {\bibinfo  {journal} {MNRAS}\ } (\bibinfo {year} {2014}),\
  10.1093/mnras/stv774},\ \Eprint {http://arxiv.org/abs/arXiv:1410.2236}
  {arXiv:1410.2236} \BibitemShut {NoStop}%
\bibitem [{\citenamefont {{Sidhu}}\ and\ \citenamefont
  {{Starkman}}(2020)}]{Sidhu2020reconsider}%
  \BibitemOpen
  \bibfield  {author} {\bibinfo {author} {\bibfnamefont {J.~S.}\ \bibnamefont
  {{Sidhu}}}\ and\ \bibinfo {author} {\bibfnamefont {G.~D.}\ \bibnamefont
  {{Starkman}}},\ }\href {\doibase 10.1103/PhysRevD.101.083503} {\bibfield
  {journal} {\bibinfo  {journal} {Phys. Rev. D}\ }\textbf {\bibinfo {volume}
  {101}},\ \bibinfo {pages} {083503} (\bibinfo {year} {2020})},\ \Eprint
  {http://arxiv.org/abs/1912.04053} {arXiv:1912.04053 [astro-ph.CO]}
  \BibitemShut {NoStop}%
\bibitem [{\citenamefont {Witten}(1984)}]{Witten1984}%
  \BibitemOpen
  \bibfield  {author} {\bibinfo {author} {\bibfnamefont {E.}~\bibnamefont
  {Witten}},\ }\href {\doibase 10.1103/physrevd.30.272} {\bibfield  {journal}
  {\bibinfo  {journal} {Physical Review D}\ }\textbf {\bibinfo {volume} {30}},\
  \bibinfo {pages} {272} (\bibinfo {year} {1984})}\BibitemShut {NoStop}%
\bibitem [{\citenamefont {Lynn}\ \emph {et~al.}(1990)\citenamefont {Lynn},
  \citenamefont {Nelson},\ and\ \citenamefont {Tetradis}}]{Lynn1990}%
  \BibitemOpen
  \bibfield  {author} {\bibinfo {author} {\bibfnamefont {B.~W.}\ \bibnamefont
  {Lynn}}, \bibinfo {author} {\bibfnamefont {A.~E.}\ \bibnamefont {Nelson}}, \
  and\ \bibinfo {author} {\bibfnamefont {N.}~\bibnamefont {Tetradis}},\
  }\href@noop {} {\bibfield  {journal} {\bibinfo  {journal} {Nuclear Physics
  B}\ }\textbf {\bibinfo {volume} {345}} (\bibinfo {year} {1990})}\BibitemShut
  {NoStop}%
\bibitem [{\citenamefont {Lynn}(2010)}]{Lynn2010}%
  \BibitemOpen
  \bibfield  {author} {\bibinfo {author} {\bibfnamefont {B.~W.}\ \bibnamefont
  {Lynn}},\ }\href@noop {} {\enquote {\bibinfo {title} {Liquid phases in
  {SU(3)} {C}hiral {P}erturbation {T}heory: {D}rops of {S}trange {C}hiral
  {N}ucleon {L}iquid and {O}rdinary {C}hiral {H}eavy {N}uclear {L}iquid},}\ }
  (\bibinfo {year} {2010}),\ \Eprint {http://arxiv.org/abs/arXiv:1005.2124}
  {arXiv:1005.2124} \BibitemShut {NoStop}%
\bibitem [{\citenamefont {Nelson}(1990)}]{Nelson1990iu}%
  \BibitemOpen
  \bibfield  {author} {\bibinfo {author} {\bibfnamefont {A.~E.}\ \bibnamefont
  {Nelson}},\ }\href {\doibase 10.1016/0370-2693(90)90429-A} {\bibfield
  {journal} {\bibinfo  {journal} {Physical Letters}\ }\textbf {\bibinfo
  {volume} {B240}},\ \bibinfo {pages} {179} (\bibinfo {year}
  {1990})}\BibitemShut {NoStop}%
\bibitem [{\citenamefont {Zhitnitsky}(2003)}]{Zhitnitsky2003}%
  \BibitemOpen
  \bibfield  {author} {\bibinfo {author} {\bibfnamefont {A.~R.}\ \bibnamefont
  {Zhitnitsky}},\ }\href {\doibase 10.1088/1475-7516/2003/10/010} {\bibfield
  {journal} {\bibinfo  {journal} {Journal of Cosmology and Astroparticle
  Physics}\ }\textbf {\bibinfo {volume} {2003}},\ \bibinfo {pages} {010}
  (\bibinfo {year} {2003})}\BibitemShut {NoStop}%
\bibitem [{\citenamefont {Jacobs}\ \emph {et~al.}(2015)\citenamefont {Jacobs},
  \citenamefont {Weltman},\ and\ \citenamefont
  {Starkman}}]{jacobs2015resonant}%
  \BibitemOpen
  \bibfield  {author} {\bibinfo {author} {\bibfnamefont {D.~M.}\ \bibnamefont
  {Jacobs}}, \bibinfo {author} {\bibfnamefont {A.}~\bibnamefont {Weltman}}, \
  and\ \bibinfo {author} {\bibfnamefont {G.~D.}\ \bibnamefont {Starkman}},\
  }\href {\doibase 10.1103/physrevd.91.115023} {\bibfield  {journal} {\bibinfo
  {journal} {Physical Review D}\ }\textbf {\bibinfo {volume} {91}},\ \bibinfo
  {pages} {115023} (\bibinfo {year} {2015})}\BibitemShut {NoStop}%
\bibitem [{\citenamefont {Sidhu}\ \emph {et~al.}(2020)\citenamefont {Sidhu},
  \citenamefont {Scherrer},\ and\ \citenamefont {Starkman}}]{Sidhu2019death}%
  \BibitemOpen
  \bibfield  {author} {\bibinfo {author} {\bibfnamefont {J.~S.}\ \bibnamefont
  {Sidhu}}, \bibinfo {author} {\bibfnamefont {R.}~\bibnamefont {Scherrer}}, \
  and\ \bibinfo {author} {\bibfnamefont {G.}~\bibnamefont {Starkman}},\ }\href
  {\doibase 10.1016/j.physletb.2020.135300} {\bibfield  {journal} {\bibinfo
  {journal} {Physics Letters B}\ }\textbf {\bibinfo {volume} {803}},\ \bibinfo
  {pages} {135300} (\bibinfo {year} {2020})}\BibitemShut {NoStop}%
\bibitem [{\citenamefont {Sidhu}\ and\ \citenamefont
  {Starkman}(2019)}]{Sidhu2019bolide}%
  \BibitemOpen
  \bibfield  {author} {\bibinfo {author} {\bibfnamefont {J.~S.}\ \bibnamefont
  {Sidhu}}\ and\ \bibinfo {author} {\bibfnamefont {G.}~\bibnamefont
  {Starkman}},\ }\href {\doibase 10.1103/PhysRevD.100.123008} {\bibfield
  {journal} {\bibinfo  {journal} {Phys. Rev. D}\ }\textbf {\bibinfo {volume}
  {100}},\ \bibinfo {pages} {123008} (\bibinfo {year} {2019})},\ \Eprint
  {http://arxiv.org/abs/1908.00557} {arXiv:1908.00557 [astro-ph.CO]}
  \BibitemShut {NoStop}%
\bibitem [{\citenamefont {Price}(1988)}]{Price1988ge}%
  \BibitemOpen
  \bibfield  {author} {\bibinfo {author} {\bibfnamefont {P.~B.}\ \bibnamefont
  {Price}},\ }\href {\doibase 10.1103/PhysRevD.38.3813} {\bibfield  {journal}
  {\bibinfo  {journal} {Physical Review D}\ }\textbf {\bibinfo {volume} {38}},\
  \bibinfo {pages} {3813} (\bibinfo {year} {1988})}\BibitemShut {NoStop}%
\bibitem [{\citenamefont {De~Rujula}\ and\ \citenamefont
  {Glashow}(1984)}]{DeRujula1984axn}%
  \BibitemOpen
  \bibfield  {author} {\bibinfo {author} {\bibfnamefont {A.}~\bibnamefont
  {De~Rujula}}\ and\ \bibinfo {author} {\bibfnamefont {S.~L.}\ \bibnamefont
  {Glashow}},\ }\href {\doibase 10.1038/312734a0} {\bibfield  {journal}
  {\bibinfo  {journal} {Nature}\ }\textbf {\bibinfo {volume} {312}},\ \bibinfo
  {pages} {734} (\bibinfo {year} {1984})}\BibitemShut {NoStop}%
\bibitem [{\citenamefont {Alcock}\ \emph {et~al.}(2001)\citenamefont {Alcock}
  \emph {et~al.}}]{Alcock2001}%
  \BibitemOpen
  \bibfield  {author} {\bibinfo {author} {\bibfnamefont {C.}~\bibnamefont
  {Alcock}} \emph {et~al.},\ }\href {\doibase 10.1086/319636} {\bibfield
  {journal} {\bibinfo  {journal} {The Astrophysical Journal}\ }\textbf
  {\bibinfo {volume} {550}},\ \bibinfo {pages} {L169} (\bibinfo {year}
  {2001})}\BibitemShut {NoStop}%
\bibitem [{\citenamefont {Griest}\ \emph {et~al.}(2013)\citenamefont {Griest},
  \citenamefont {Cieplak},\ and\ \citenamefont {Lehner}}]{Griest2013}%
  \BibitemOpen
  \bibfield  {author} {\bibinfo {author} {\bibfnamefont {K.}~\bibnamefont
  {Griest}}, \bibinfo {author} {\bibfnamefont {A.~M.}\ \bibnamefont {Cieplak}},
  \ and\ \bibinfo {author} {\bibfnamefont {M.~J.}\ \bibnamefont {Lehner}},\
  }\href {\doibase 10.1103/physrevlett.111.181302} {\bibfield  {journal}
  {\bibinfo  {journal} {Physical Review Letters}\ }\textbf {\bibinfo {volume}
  {111}},\ \bibinfo {pages} {181302} (\bibinfo {year} {2013})}\BibitemShut
  {NoStop}%
\bibitem [{\citenamefont {Tisserand}\ \emph {et~al.}(2007)\citenamefont
  {Tisserand} \emph {et~al.}}]{Tisserand2007}%
  \BibitemOpen
  \bibfield  {author} {\bibinfo {author} {\bibfnamefont {P.}~\bibnamefont
  {Tisserand}} \emph {et~al.},\ }\href {\doibase 10.1051/0004-6361:20066017}
  {\bibfield  {journal} {\bibinfo  {journal} {Astronomy {\&} Astrophysics}\
  }\textbf {\bibinfo {volume} {469}},\ \bibinfo {pages} {387} (\bibinfo {year}
  {2007})}\BibitemShut {NoStop}%
\bibitem [{\citenamefont {Carr}\ \emph {et~al.}(2010)\citenamefont {Carr},
  \citenamefont {Kohri}, \citenamefont {Sendouda},\ and\ \citenamefont
  {Yokoyama}}]{Carr2010}%
  \BibitemOpen
  \bibfield  {author} {\bibinfo {author} {\bibfnamefont {B.~J.}\ \bibnamefont
  {Carr}}, \bibinfo {author} {\bibfnamefont {K.}~\bibnamefont {Kohri}},
  \bibinfo {author} {\bibfnamefont {Y.}~\bibnamefont {Sendouda}}, \ and\
  \bibinfo {author} {\bibfnamefont {J.}~\bibnamefont {Yokoyama}},\ }\href
  {\doibase 10.1103/physrevd.81.104019} {\bibfield  {journal} {\bibinfo
  {journal} {Physical Review D}\ }\textbf {\bibinfo {volume} {81}},\ \bibinfo
  {pages} {104019} (\bibinfo {year} {2010})}\BibitemShut {NoStop}%
\bibitem [{\citenamefont {Niikura}\ \emph {et~al.}(2019)\citenamefont {Niikura}
  \emph {et~al.}}]{Niikura2019}%
  \BibitemOpen
  \bibfield  {author} {\bibinfo {author} {\bibfnamefont {H.}~\bibnamefont
  {Niikura}} \emph {et~al.},\ }\href {\doibase 10.1038/s41550-019-0723-1}
  {\bibfield  {journal} {\bibinfo  {journal} {Nature Astronomy}\ }\textbf
  {\bibinfo {volume} {3}},\ \bibinfo {pages} {524} (\bibinfo {year}
  {2019})}\BibitemShut {NoStop}%
\bibitem [{\citenamefont {Wilkinson}\ \emph {et~al.}(2014)\citenamefont
  {Wilkinson}, \citenamefont {Lesgourgues},\ and\ \citenamefont
  {B{\oe}hm}}]{Wilkinson2014angular}%
  \BibitemOpen
  \bibfield  {author} {\bibinfo {author} {\bibfnamefont {R.~J.}\ \bibnamefont
  {Wilkinson}}, \bibinfo {author} {\bibfnamefont {J.}~\bibnamefont
  {Lesgourgues}}, \ and\ \bibinfo {author} {\bibfnamefont {C.}~\bibnamefont
  {B{\oe}hm}},\ }\href {\doibase 10.1088/1475-7516/2014/04/026} {\bibfield
  {journal} {\bibinfo  {journal} {Journal of Cosmology and Astroparticle
  Physics}\ }\textbf {\bibinfo {volume} {2014}},\ \bibinfo {pages} {026}
  (\bibinfo {year} {2014})}\BibitemShut {NoStop}%
\bibitem [{\citenamefont {Graham}\ \emph {et~al.}(2018)\citenamefont {Graham},
  \citenamefont {Janish}, \citenamefont {Narayan}, \citenamefont {Rajendran},\
  and\ \citenamefont {Riggins}}]{Graham2018}%
  \BibitemOpen
  \bibfield  {author} {\bibinfo {author} {\bibfnamefont {P.~W.}\ \bibnamefont
  {Graham}}, \bibinfo {author} {\bibfnamefont {R.}~\bibnamefont {Janish}},
  \bibinfo {author} {\bibfnamefont {V.}~\bibnamefont {Narayan}}, \bibinfo
  {author} {\bibfnamefont {S.}~\bibnamefont {Rajendran}}, \ and\ \bibinfo
  {author} {\bibfnamefont {P.}~\bibnamefont {Riggins}},\ }\href {\doibase
  10.1103/physrevd.98.115027} {\bibfield  {journal} {\bibinfo  {journal}
  {Physical Review D}\ }\textbf {\bibinfo {volume} {98}},\ \bibinfo {pages}
  {115027} (\bibinfo {year} {2018})}\BibitemShut {NoStop}%
\bibitem [{\citenamefont {Sidhu}(2020)}]{Sidhu2020charge}%
  \BibitemOpen
  \bibfield  {author} {\bibinfo {author} {\bibfnamefont {J.~S.}\ \bibnamefont
  {Sidhu}},\ }\href {\doibase 10.1103/PhysRevD.101.043526} {\bibfield
  {journal} {\bibinfo  {journal} {Phys. Rev. D}\ }\textbf {\bibinfo {volume}
  {101}},\ \bibinfo {pages} {043526} (\bibinfo {year} {2020})},\ \Eprint
  {http://arxiv.org/abs/1912.04732} {arXiv:1912.04732 [astro-ph.CO]}
  \BibitemShut {NoStop}%
\bibitem [{\citenamefont {{Sidhu}}\ \emph {et~al.}(2020)\citenamefont
  {{Sidhu}}, \citenamefont {{Scherrer}},\ and\ \citenamefont
  {{Starkman}}}]{Sidhu2020anti}%
  \BibitemOpen
  \bibfield  {author} {\bibinfo {author} {\bibfnamefont {J.~S.}\ \bibnamefont
  {{Sidhu}}}, \bibinfo {author} {\bibfnamefont {R.~J.}\ \bibnamefont
  {{Scherrer}}}, \ and\ \bibinfo {author} {\bibfnamefont {G.~D.}\ \bibnamefont
  {{Starkman}}},\ }\href@noop {} {\enquote {\bibinfo {title} {Antimatter as
  macroscopic dark matter},}\ } (\bibinfo {year} {2020}),\ \Eprint
  {http://arxiv.org/abs/arXiv:2006.01200} {arXiv:2006.01200} \BibitemShut
  {NoStop}%
\bibitem [{\citenamefont {Sidhu}\ \emph
  {et~al.}(2019{\natexlab{a}})\citenamefont {Sidhu}, \citenamefont {Abraham},
  \citenamefont {Covault},\ and\ \citenamefont {Starkman}}]{Sidhu2018auv}%
  \BibitemOpen
  \bibfield  {author} {\bibinfo {author} {\bibfnamefont {J.~S.}\ \bibnamefont
  {Sidhu}}, \bibinfo {author} {\bibfnamefont {R.~M.}\ \bibnamefont {Abraham}},
  \bibinfo {author} {\bibfnamefont {C.}~\bibnamefont {Covault}}, \ and\
  \bibinfo {author} {\bibfnamefont {G.}~\bibnamefont {Starkman}},\ }\href
  {\doibase 10.1088/1475-7516/2019/02/037} {\bibfield  {journal} {\bibinfo
  {journal} {JCAP}\ }\textbf {\bibinfo {volume} {1902}},\ \bibinfo {pages}
  {037} (\bibinfo {year} {2019}{\natexlab{a}})},\ \Eprint
  {http://arxiv.org/abs/1808.06978} {1808.06978 [astro-ph.HE]} \BibitemShut
  {NoStop}%
\bibitem [{\citenamefont {Abraham}\ \emph {et~al.}(2010)\citenamefont {Abraham}
  \emph {et~al.}}]{Abraham2010}%
  \BibitemOpen
  \bibfield  {author} {\bibinfo {author} {\bibfnamefont {J.}~\bibnamefont
  {Abraham}} \emph {et~al.},\ }\href {\doibase 10.1016/j.nima.2010.04.023}
  {\bibfield  {journal} {\bibinfo  {journal} {Nuclear Instruments and Methods
  in Physics Research Section A: Accelerators, Spectrometers, Detectors and
  Associated Equipment}\ }\textbf {\bibinfo {volume} {620}},\ \bibinfo {pages}
  {227} (\bibinfo {year} {2010})}\BibitemShut {NoStop}%
\bibitem [{\citenamefont {Sidhu}\ \emph
  {et~al.}(2019{\natexlab{b}})\citenamefont {Sidhu}, \citenamefont {Starkman},\
  and\ \citenamefont {Harvey}}]{Sidhu2019granite}%
  \BibitemOpen
  \bibfield  {author} {\bibinfo {author} {\bibfnamefont {J.~S.}\ \bibnamefont
  {Sidhu}}, \bibinfo {author} {\bibfnamefont {G.}~\bibnamefont {Starkman}}, \
  and\ \bibinfo {author} {\bibfnamefont {R.}~\bibnamefont {Harvey}},\ }\href
  {\doibase 10.1103/PhysRevD.100.103015} {\bibfield  {journal} {\bibinfo
  {journal} {Phys. Rev. D}\ }\textbf {\bibinfo {volume} {100}},\ \bibinfo
  {pages} {103015} (\bibinfo {year} {2019}{\natexlab{b}})},\ \Eprint
  {http://arxiv.org/abs/1905.10025} {arXiv:1905.10025 [astro-ph.HE]}
  \BibitemShut {NoStop}%
\bibitem [{\citenamefont {Dwyer}\ and\ \citenamefont
  {Uman}(2014)}]{DwyerUman2014}%
  \BibitemOpen
  \bibfield  {author} {\bibinfo {author} {\bibfnamefont {J.~R.}\ \bibnamefont
  {Dwyer}}\ and\ \bibinfo {author} {\bibfnamefont {M.~A.}\ \bibnamefont
  {Uman}},\ }\href@noop {} {\bibfield  {journal} {\bibinfo  {journal} {Physics
  Reports}\ }\textbf {\bibinfo {volume} {534}},\ \bibinfo {pages} {147}
  (\bibinfo {year} {2014})}\BibitemShut {NoStop}%
\bibitem [{\citenamefont {Babich}\ \emph {et~al.}(2012)\citenamefont {Babich},
  \citenamefont {Bochkov}, \citenamefont {Dwyer},\ and\ \citenamefont
  {Kutsyk}}]{Babich2012}%
  \BibitemOpen
  \bibfield  {author} {\bibinfo {author} {\bibfnamefont {L.~P.}\ \bibnamefont
  {Babich}}, \bibinfo {author} {\bibfnamefont {E.~I.}\ \bibnamefont {Bochkov}},
  \bibinfo {author} {\bibfnamefont {J.~R.}\ \bibnamefont {Dwyer}}, \ and\
  \bibinfo {author} {\bibfnamefont {I.~M.}\ \bibnamefont {Kutsyk}},\ }\href
  {\doibase 10.1029/2012JA017799} {\bibfield  {journal} {\bibinfo  {journal}
  {Journal of Geophysical Research: Space Physics}\ }\textbf {\bibinfo {volume}
  {117}} (\bibinfo {year} {2012}),\ 10.1029/2012JA017799}\BibitemShut {NoStop}%
\bibitem [{\citenamefont {American}(2008)}]{scientific_american_2008}%
  \BibitemOpen
  \bibfield  {author} {\bibinfo {author} {\bibfnamefont {S.}~\bibnamefont
  {American}},\ }\href
  {https://www.scientificamerican.com/article/experts-do-cosmic-rays-cause-lightning/}
  {\bibfield  {journal} {\bibinfo  {journal} {Scientific American}\ } (\bibinfo
  {year} {2008})}\BibitemShut {NoStop}%
\bibitem [{\citenamefont {Hill}\ \emph {et~al.}(2012)\citenamefont {Hill},
  \citenamefont {Pilkey}, \citenamefont {Uman}, \citenamefont {Jordan},
  \citenamefont {Rison},\ and\ \citenamefont {Krehbiel}}]{Hill2012}%
  \BibitemOpen
  \bibfield  {author} {\bibinfo {author} {\bibfnamefont {J.~D.}\ \bibnamefont
  {Hill}}, \bibinfo {author} {\bibfnamefont {J.}~\bibnamefont {Pilkey}},
  \bibinfo {author} {\bibfnamefont {M.~A.}\ \bibnamefont {Uman}}, \bibinfo
  {author} {\bibfnamefont {D.~M.}\ \bibnamefont {Jordan}}, \bibinfo {author}
  {\bibfnamefont {W.}~\bibnamefont {Rison}}, \ and\ \bibinfo {author}
  {\bibfnamefont {P.~R.}\ \bibnamefont {Krehbiel}},\ }\href
  {https://doi.org/10.1029/2012gl051932} {\bibfield  {journal} {\bibinfo
  {journal} {Geophysical Research Letters}\ }\textbf {\bibinfo {volume} {39}},\
  \bibinfo {pages} {n/a} (\bibinfo {year} {2012})}\BibitemShut {NoStop}%
\bibitem [{\citenamefont {Hill}\ \emph {et~al.}(2013)\citenamefont {Hill},
  \citenamefont {Pilkey}, \citenamefont {Uman}, \citenamefont {Jordan},
  \citenamefont {Rison}, \citenamefont {Krebhiel}, \citenamefont {Biggerstaff},
  \citenamefont {Hyland},\ and\ \citenamefont {Blakeslee}}]{Hill2013}%
  \BibitemOpen
  \bibfield  {author} {\bibinfo {author} {\bibfnamefont {J.~D.}\ \bibnamefont
  {Hill}}, \bibinfo {author} {\bibfnamefont {J.}~\bibnamefont {Pilkey}},
  \bibinfo {author} {\bibfnamefont {M.~A.}\ \bibnamefont {Uman}}, \bibinfo
  {author} {\bibfnamefont {D.~M.}\ \bibnamefont {Jordan}}, \bibinfo {author}
  {\bibfnamefont {W.}~\bibnamefont {Rison}}, \bibinfo {author} {\bibfnamefont
  {P.~R.}\ \bibnamefont {Krebhiel}}, \bibinfo {author} {\bibfnamefont {M.~I.}\
  \bibnamefont {Biggerstaff}}, \bibinfo {author} {\bibfnamefont
  {P.}~\bibnamefont {Hyland}}, \ and\ \bibinfo {author} {\bibfnamefont
  {R.}~\bibnamefont {Blakeslee}},\ }\href {\doibase 10.1002/jgrd.50660}
  {\bibfield  {journal} {\bibinfo  {journal} {Journal of Geophysical Research:
  Atmospheres}\ }\textbf {\bibinfo {volume} {118}},\ \bibinfo {pages} {8460}
  (\bibinfo {year} {2013})}\BibitemShut {NoStop}%
\bibitem [{\citenamefont {Wang}\ \emph {et~al.}(1999)\citenamefont {Wang},
  \citenamefont {Rakov}, \citenamefont {Uman}, \citenamefont {Takagi},
  \citenamefont {Watanabe}, \citenamefont {Crawford}, \citenamefont {Rambo},
  \citenamefont {Schnetzer}, \citenamefont {Fisher},\ and\ \citenamefont
  {Kawasaki}}]{Wang1999}%
  \BibitemOpen
  \bibfield  {author} {\bibinfo {author} {\bibfnamefont {D.}~\bibnamefont
  {Wang}}, \bibinfo {author} {\bibfnamefont {V.~A.}\ \bibnamefont {Rakov}},
  \bibinfo {author} {\bibfnamefont {M.~A.}\ \bibnamefont {Uman}}, \bibinfo
  {author} {\bibfnamefont {N.}~\bibnamefont {Takagi}}, \bibinfo {author}
  {\bibfnamefont {T.}~\bibnamefont {Watanabe}}, \bibinfo {author}
  {\bibfnamefont {D.~E.}\ \bibnamefont {Crawford}}, \bibinfo {author}
  {\bibfnamefont {K.~J.}\ \bibnamefont {Rambo}}, \bibinfo {author}
  {\bibfnamefont {G.~H.}\ \bibnamefont {Schnetzer}}, \bibinfo {author}
  {\bibfnamefont {R.~J.}\ \bibnamefont {Fisher}}, \ and\ \bibinfo {author}
  {\bibfnamefont {Z.-I.}\ \bibnamefont {Kawasaki}},\ }\href {\doibase
  10.1029/1998jd200070} {\bibfield  {journal} {\bibinfo  {journal} {Journal of
  Geophysical Research: Atmospheres}\ }\textbf {\bibinfo {volume} {104}},\
  \bibinfo {pages} {2143} (\bibinfo {year} {1999})}\BibitemShut {NoStop}%
\bibitem [{\citenamefont {Pilkey}\ \emph {et~al.}(2013)\citenamefont {Pilkey},
  \citenamefont {Uman}, \citenamefont {Hill}, \citenamefont {Ngin},
  \citenamefont {Gamerota}, \citenamefont {Jordan}, \citenamefont {Rison},
  \citenamefont {Krehbiel}, \citenamefont {Edens}, \citenamefont
  {Biggerstaff},\ and\ \citenamefont {Hyland}}]{rocket2012}%
  \BibitemOpen
  \bibfield  {author} {\bibinfo {author} {\bibfnamefont {J.~T.}\ \bibnamefont
  {Pilkey}}, \bibinfo {author} {\bibfnamefont {M.~A.}\ \bibnamefont {Uman}},
  \bibinfo {author} {\bibfnamefont {J.~D.}\ \bibnamefont {Hill}}, \bibinfo
  {author} {\bibfnamefont {T.}~\bibnamefont {Ngin}}, \bibinfo {author}
  {\bibfnamefont {W.~R.}\ \bibnamefont {Gamerota}}, \bibinfo {author}
  {\bibfnamefont {D.~M.}\ \bibnamefont {Jordan}}, \bibinfo {author}
  {\bibfnamefont {W.}~\bibnamefont {Rison}}, \bibinfo {author} {\bibfnamefont
  {P.~R.}\ \bibnamefont {Krehbiel}}, \bibinfo {author} {\bibfnamefont {H.~E.}\
  \bibnamefont {Edens}}, \bibinfo {author} {\bibfnamefont {M.~I.}\ \bibnamefont
  {Biggerstaff}}, \ and\ \bibinfo {author} {\bibfnamefont {P.}~\bibnamefont
  {Hyland}},\ }\href {\doibase 10.1002/2013JD020501} {\bibfield  {journal}
  {\bibinfo  {journal} {Journal of Geophysical Research: Atmospheres}\ }\textbf
  {\bibinfo {volume} {118}},\ \bibinfo {pages} {13,158} (\bibinfo {year}
  {2013})}\BibitemShut {NoStop}%
\bibitem [{\citenamefont {Cyncynates}\ \emph {et~al.}(2017)\citenamefont
  {Cyncynates}, \citenamefont {Chiel}, \citenamefont {Sidhu},\ and\
  \citenamefont {Starkman}}]{Cyncynates2016}%
  \BibitemOpen
  \bibfield  {author} {\bibinfo {author} {\bibfnamefont {D.}~\bibnamefont
  {Cyncynates}}, \bibinfo {author} {\bibfnamefont {J.}~\bibnamefont {Chiel}},
  \bibinfo {author} {\bibfnamefont {J.}~\bibnamefont {Sidhu}}, \ and\ \bibinfo
  {author} {\bibfnamefont {G.~D.}\ \bibnamefont {Starkman}},\ }\href {\doibase
  10.1103/PhysRevD.95.063006} {\bibfield  {journal} {\bibinfo  {journal} {Phys.
  Rev. D}\ }\textbf {\bibinfo {volume} {95}},\ \bibinfo {pages} {063006}
  (\bibinfo {year} {2017})},\ \bibinfo {note} {[Addendum: Phys.Rev.D 95, 129903
  (2017)]},\ \Eprint {http://arxiv.org/abs/1610.09680} {arXiv:1610.09680
  [astro-ph.CO]} \BibitemShut {NoStop}%
\bibitem [{\citenamefont {Picone}\ and\ \citenamefont
  {Boris}(1983)}]{Picone1983}%
  \BibitemOpen
  \bibfield  {author} {\bibinfo {author} {\bibfnamefont {J.~M.}\ \bibnamefont
  {Picone}}\ and\ \bibinfo {author} {\bibfnamefont {J.~P.}\ \bibnamefont
  {Boris}},\ }\href {\doibase 10.1063/1.864173} {\bibfield  {journal} {\bibinfo
   {journal} {The Physics of Fluids}\ }\textbf {\bibinfo {volume} {26}},\
  \bibinfo {pages} {365} (\bibinfo {year} {1983})}\BibitemShut {NoStop}%
\bibitem [{\citenamefont {Capitelli}\ \emph {et~al.}(2000)\citenamefont
  {Capitelli}, \citenamefont {Colonna}, \citenamefont {Gorse},\ and\
  \citenamefont {D'Angola}}]{Capitelli2000}%
  \BibitemOpen
  \bibfield  {author} {\bibinfo {author} {\bibfnamefont {M.}~\bibnamefont
  {Capitelli}}, \bibinfo {author} {\bibfnamefont {G.}~\bibnamefont {Colonna}},
  \bibinfo {author} {\bibfnamefont {C.}~\bibnamefont {Gorse}}, \ and\ \bibinfo
  {author} {\bibfnamefont {A.}~\bibnamefont {D'Angola}},\ }\href@noop {}
  {\bibfield  {journal} {\bibinfo  {journal} {Euro. Phys. Jour. D}\ }\textbf
  {\bibinfo {volume} {11}},\ \bibinfo {pages} {278} (\bibinfo {year}
  {2000})}\BibitemShut {NoStop}%
\bibitem [{\citenamefont {Eisazadeh-Far}\ \emph {et~al.}(2011)\citenamefont
  {Eisazadeh-Far}, \citenamefont {Metghalchi},\ and\ \citenamefont
  {Keck}}]{EisazadehFar2011}%
  \BibitemOpen
  \bibfield  {author} {\bibinfo {author} {\bibfnamefont {K.}~\bibnamefont
  {Eisazadeh-Far}}, \bibinfo {author} {\bibfnamefont {H.}~\bibnamefont
  {Metghalchi}}, \ and\ \bibinfo {author} {\bibfnamefont {J.~C.}\ \bibnamefont
  {Keck}},\ }\href {\doibase 10.1115/1.4003881} {\bibfield  {journal} {\bibinfo
   {journal} {Journal of Energy Resources Technology}\ }\textbf {\bibinfo
  {volume} {133}} (\bibinfo {year} {2011}),\ 10.1115/1.4003881}\BibitemShut
  {NoStop}%
\bibitem [{\citenamefont {Idone}\ \emph {et~al.}(1987)\citenamefont {Idone},
  \citenamefont {Orville}, \citenamefont {Mach},\ and\ \citenamefont
  {Rust}}]{Idone1987}%
  \BibitemOpen
  \bibfield  {author} {\bibinfo {author} {\bibfnamefont {V.~P.}\ \bibnamefont
  {Idone}}, \bibinfo {author} {\bibfnamefont {R.~E.}\ \bibnamefont {Orville}},
  \bibinfo {author} {\bibfnamefont {D.~M.}\ \bibnamefont {Mach}}, \ and\
  \bibinfo {author} {\bibfnamefont {W.~D.}\ \bibnamefont {Rust}},\ }\href
  {\doibase 10.1029/gl014i011p01150} {\enquote {\bibinfo {title} {{The
  propagation speed of a positive lightning return stroke}},}\ } (\bibinfo
  {year} {1987})\BibitemShut {NoStop}%
\bibitem [{\citenamefont {Betts}(2003)}]{Betts2003}%
  \BibitemOpen
  \bibfield  {author} {\bibinfo {author} {\bibfnamefont {R.}~\bibnamefont
  {Betts}},\ }\href {https://patents.google.com/patent/US6597559B2/en}
  {\enquote {\bibinfo {title} {Lightning rocket},}\ } (\bibinfo {year}
  {2003}),\ \bibinfo {note} {{US Patent 6597559B2}}\BibitemShut {NoStop}%
\bibitem [{\citenamefont {{Hare}}(2020)}]{Hare2020}%
  \BibitemOpen
  \bibfield  {author} {\bibinfo {author} {\bibfnamefont {e.}~\bibnamefont
  {{Hare}}, \bibfnamefont {B.~M.}},\ }\href {\doibase
  10.1103/PhysRevLett.124.105101} {\bibfield  {journal} {\bibinfo  {journal}
  {Physical Review Letters}\ }\textbf {\bibinfo {volume} {124}},\ \bibinfo
  {eid} {105101} (\bibinfo {year} {2020})},\ \Eprint
  {http://arxiv.org/abs/2007.03231} {arXiv:2007.03231 [physics.ao-ph]}
  \BibitemShut {NoStop}%
\bibitem [{Note1()}]{Note1}%
  \BibitemOpen
  \bibinfo {note} {This is the distribution of macro velocities in a
  non-orbiting frame moving with the Galaxy. When considering the velocity of
  macros impacting the atmosphere, \protect \textup {\hbox {\mathsurround \z@
  \protect \normalfont (\ignorespaces \ref {eq:maxwellian}\unskip \@@italiccorr
  )}} is modified by the motion of the Sun and Earth in that frame, and by the
  Sun's and Earth's gravitational potential. We have taken into account these
  effects (as explained in \protect \citet {Freese2013}), except the negligible
  effect of Earth's gravitational potential. \par Recent hydrodynamical
  simulations of Milky Way-like galaxies including baryons, which have a
  non-negligible effect on the dark matter distribution in the Solar
  neighborhood \protect \citep {Tanabashi2018} have been performed to determine
  the correctness of assuming a Maxwellian distribution. These simulations find
  that the velocity distributions are indeed close to Maxwellian. As discussed
  previously, macros are expected to move according to \protect \textup {\hbox
  {\mathsurround \z@ \protect \normalfont (\ignorespaces \ref
  {eq:maxwellian}\unskip \@@italiccorr )}}. Taking this minimum speed
  requirement into account, we find that $71\%$ of all macros in the
  distribution will be moving at at least $250\protect \tmspace +\thinmuskip
  {.1667em}$km/s.}\BibitemShut {Stop}%
\bibitem [{\citenamefont {Kasparian}\ \emph {et~al.}(2008)\citenamefont
  {Kasparian}, \citenamefont {Ackermann}, \citenamefont {Andr\'{e}},
  \citenamefont {M\'{e}chain}, \citenamefont {M\'{e}jean}, \citenamefont
  {Prade}, \citenamefont {Rohwetter}, \citenamefont {Salmon}, \citenamefont
  {Stelmaszczyk}, \citenamefont {Yu}, \citenamefont {Mysyrowicz}, \citenamefont
  {Sauerbrey}, \citenamefont {W\"{o}ste},\ and\ \citenamefont
  {Wolf}}]{Kasparian2008}%
  \BibitemOpen
  \bibfield  {author} {\bibinfo {author} {\bibfnamefont {J.}~\bibnamefont
  {Kasparian}}, \bibinfo {author} {\bibfnamefont {R.}~\bibnamefont
  {Ackermann}}, \bibinfo {author} {\bibfnamefont {Y.-B.}\ \bibnamefont
  {Andr\'{e}}}, \bibinfo {author} {\bibfnamefont {G.}~\bibnamefont
  {M\'{e}chain}}, \bibinfo {author} {\bibfnamefont {G.}~\bibnamefont
  {M\'{e}jean}}, \bibinfo {author} {\bibfnamefont {B.}~\bibnamefont {Prade}},
  \bibinfo {author} {\bibfnamefont {P.}~\bibnamefont {Rohwetter}}, \bibinfo
  {author} {\bibfnamefont {E.}~\bibnamefont {Salmon}}, \bibinfo {author}
  {\bibfnamefont {K.}~\bibnamefont {Stelmaszczyk}}, \bibinfo {author}
  {\bibfnamefont {J.}~\bibnamefont {Yu}}, \bibinfo {author} {\bibfnamefont
  {A.}~\bibnamefont {Mysyrowicz}}, \bibinfo {author} {\bibfnamefont
  {R.}~\bibnamefont {Sauerbrey}}, \bibinfo {author} {\bibfnamefont
  {L.}~\bibnamefont {W\"{o}ste}}, \ and\ \bibinfo {author} {\bibfnamefont
  {J.-P.}\ \bibnamefont {Wolf}},\ }\href {\doibase 10.1364/OE.16.005757}
  {\bibfield  {journal} {\bibinfo  {journal} {Opt. Express}\ }\textbf {\bibinfo
  {volume} {16}},\ \bibinfo {pages} {5757} (\bibinfo {year}
  {2008})}\BibitemShut {NoStop}%
\bibitem [{\citenamefont {Colvin}\ \emph {et~al.}(1987)\citenamefont {Colvin},
  \citenamefont {Mitchell}, \citenamefont {Greig}, \citenamefont {Murphy},
  \citenamefont {Pechacek},\ and\ \citenamefont {Raleigh}}]{IVY-MIKE1987}%
  \BibitemOpen
  \bibfield  {author} {\bibinfo {author} {\bibfnamefont {J.~D.}\ \bibnamefont
  {Colvin}}, \bibinfo {author} {\bibfnamefont {C.~K.}\ \bibnamefont
  {Mitchell}}, \bibinfo {author} {\bibfnamefont {J.~R.}\ \bibnamefont {Greig}},
  \bibinfo {author} {\bibfnamefont {D.~P.}\ \bibnamefont {Murphy}}, \bibinfo
  {author} {\bibfnamefont {R.~E.}\ \bibnamefont {Pechacek}}, \ and\ \bibinfo
  {author} {\bibfnamefont {M.}~\bibnamefont {Raleigh}},\ }\href {\doibase
  10.1029/JD092iD05p05696} {\bibfield  {journal} {\bibinfo  {journal} {Journal
  of Geophysical Research: Atmospheres}\ }\textbf {\bibinfo {volume} {92}},\
  \bibinfo {pages} {5696} (\bibinfo {year} {1987})}\BibitemShut {NoStop}%
\bibitem [{\citenamefont {Choi}(2004)}]{Choi2004}%
  \BibitemOpen
  \bibfield  {author} {\bibinfo {author} {\bibfnamefont {E.~C.}\ \bibnamefont
  {Choi}},\ }\href {\doibase https://doi.org/10.1016/j.jweia.2003.12.001}
  {\bibfield  {journal} {\bibinfo  {journal} {Journal of Wind Engineering and
  Industrial Aerodynamics}\ }\textbf {\bibinfo {volume} {92}},\ \bibinfo
  {pages} {275 } (\bibinfo {year} {2004})}\BibitemShut {NoStop}%
\bibitem [{\citenamefont {Freese}\ \emph {et~al.}(2013)\citenamefont {Freese},
  \citenamefont {Lisanti},\ and\ \citenamefont {Savage}}]{Freese2013}%
  \BibitemOpen
  \bibfield  {author} {\bibinfo {author} {\bibfnamefont {K.}~\bibnamefont
  {Freese}}, \bibinfo {author} {\bibfnamefont {M.}~\bibnamefont {Lisanti}}, \
  and\ \bibinfo {author} {\bibfnamefont {C.}~\bibnamefont {Savage}},\ }\href
  {\doibase 10.1103/revmodphys.85.1561} {\bibfield  {journal} {\bibinfo
  {journal} {Reviews of Modern Physics}\ }\textbf {\bibinfo {volume} {85}},\
  \bibinfo {pages} {1561} (\bibinfo {year} {2013})}\BibitemShut {NoStop}%
\bibitem [{\citenamefont {Bovy}\ and\ \citenamefont
  {Tremaine}(2012)}]{Bovy2012}%
  \BibitemOpen
  \bibfield  {author} {\bibinfo {author} {\bibfnamefont {J.}~\bibnamefont
  {Bovy}}\ and\ \bibinfo {author} {\bibfnamefont {S.}~\bibnamefont
  {Tremaine}},\ }\href {\doibase 10.1088/0004-637x/756/1/89} {\bibfield
  {journal} {\bibinfo  {journal} {The Astrophysical Journal}\ }\textbf
  {\bibinfo {volume} {756}},\ \bibinfo {pages} {89} (\bibinfo {year}
  {2012})}\BibitemShut {NoStop}%
\bibitem [{\citenamefont {Dziewonski}\ and\ \citenamefont
  {Anderson}(1981)}]{Dziewonski1981}%
  \BibitemOpen
  \bibfield  {author} {\bibinfo {author} {\bibfnamefont {A.~M.}\ \bibnamefont
  {Dziewonski}}\ and\ \bibinfo {author} {\bibfnamefont {D.~L.}\ \bibnamefont
  {Anderson}},\ }\href {\doibase 10.1016/0031-9201(81)90046-7} {\bibfield
  {journal} {\bibinfo  {journal} {Physics of the Earth and Planetary
  Interiors}\ }\textbf {\bibinfo {volume} {25}},\ \bibinfo {pages} {297}
  (\bibinfo {year} {1981})}\BibitemShut {NoStop}%
\bibitem [{\citenamefont {Lowenstein}(2020)}]{Zimbabwe}%
  \BibitemOpen
  \bibfield  {author} {\bibinfo {author} {\bibfnamefont {P.}~\bibnamefont
  {Lowenstein}},\ }\href@noop {} {\bibfield  {journal} {\bibinfo  {journal}
  {Rare straight lightning over Zimbabwe,
  https://earthsky.org/todays-image/rare-straight-lightning-over-zimbabwe}\ }
  (\bibinfo {year} {Accessed: 06.13.2020})}\BibitemShut {NoStop}%
\bibitem [{\citenamefont {Downs}(2005)}]{NatGeo}%
  \BibitemOpen
  \bibfield  {author} {\bibinfo {author} {\bibfnamefont {R.}~\bibnamefont
  {Downs}},\ }\href@noop {} {\emph {\bibinfo {title} {National Geographic
  almanac of geography}}}\ (\bibinfo  {publisher} {National Geographic},\
  \bibinfo {address} {Washington, D.C},\ \bibinfo {year} {2005})\BibitemShut
  {NoStop}%
\bibitem [{\citenamefont {Christian}\ \emph {et~al.}(2003)\citenamefont
  {Christian}, \citenamefont {Blakeslee}, \citenamefont {Boccippio},
  \citenamefont {Boeck}, \citenamefont {Buechler}, \citenamefont {Driscoll},
  \citenamefont {Goodman}, \citenamefont {Hall}, \citenamefont {Koshak},
  \citenamefont {Mach},\ and\ \citenamefont {Stewart}}]{Christian2003}%
  \BibitemOpen
  \bibfield  {author} {\bibinfo {author} {\bibfnamefont {H.~J.}\ \bibnamefont
  {Christian}}, \bibinfo {author} {\bibfnamefont {R.~J.}\ \bibnamefont
  {Blakeslee}}, \bibinfo {author} {\bibfnamefont {D.~J.}\ \bibnamefont
  {Boccippio}}, \bibinfo {author} {\bibfnamefont {W.~L.}\ \bibnamefont
  {Boeck}}, \bibinfo {author} {\bibfnamefont {D.~E.}\ \bibnamefont {Buechler}},
  \bibinfo {author} {\bibfnamefont {K.~T.}\ \bibnamefont {Driscoll}}, \bibinfo
  {author} {\bibfnamefont {S.~J.}\ \bibnamefont {Goodman}}, \bibinfo {author}
  {\bibfnamefont {J.~M.}\ \bibnamefont {Hall}}, \bibinfo {author}
  {\bibfnamefont {W.~J.}\ \bibnamefont {Koshak}}, \bibinfo {author}
  {\bibfnamefont {D.~M.}\ \bibnamefont {Mach}}, \ and\ \bibinfo {author}
  {\bibfnamefont {M.~F.}\ \bibnamefont {Stewart}},\ }\href {\doibase
  10.1029/2002JD002347} {\bibfield  {journal} {\bibinfo  {journal} {Journal of
  Geophysical Research: Atmospheres}\ }\textbf {\bibinfo {volume} {108}},\
  \bibinfo {pages} {ACL 4} (\bibinfo {year} {2003})}\BibitemShut {NoStop}%
\bibitem [{\citenamefont {Mackerras}\ \emph {et~al.}(1998)\citenamefont
  {Mackerras}, \citenamefont {Darveniza}, \citenamefont {Orville},
  \citenamefont {Williams},\ and\ \citenamefont {Goodman}}]{Mackerras1998}%
  \BibitemOpen
  \bibfield  {author} {\bibinfo {author} {\bibfnamefont {D.}~\bibnamefont
  {Mackerras}}, \bibinfo {author} {\bibfnamefont {M.}~\bibnamefont
  {Darveniza}}, \bibinfo {author} {\bibfnamefont {R.~E.}\ \bibnamefont
  {Orville}}, \bibinfo {author} {\bibfnamefont {E.~R.}\ \bibnamefont
  {Williams}}, \ and\ \bibinfo {author} {\bibfnamefont {S.~J.}\ \bibnamefont
  {Goodman}},\ }\href {\doibase 10.1029/98JD01461} {\bibfield  {journal}
  {\bibinfo  {journal} {Journal of Geophysical Research: Atmospheres}\ }\textbf
  {\bibinfo {volume} {103}},\ \bibinfo {pages} {19791} (\bibinfo {year}
  {1998})}\BibitemShut {NoStop}%
\bibitem [{\citenamefont {{Brown}}\ \emph {et~al.}(2018)\citenamefont
  {{Brown}}, \citenamefont {{Janssen}}, \citenamefont {{Adumitroaie}},
  \citenamefont {{Atreya}}, \citenamefont {{Bolton}}, \citenamefont {{Gulkis}},
  \citenamefont {{Ingersoll}}, \citenamefont {{Levin}}, \citenamefont {{Li}},
  \citenamefont {{Li}}, \citenamefont {{Lunine}}, \citenamefont {{Misra}},
  \citenamefont {{Orton}}, \citenamefont {{Steffes}}, \citenamefont
  {{Tabataba-Vakili}}, \citenamefont {{Kolma{\v{s}}ov{\'a}}}, \citenamefont
  {{Imai}}, \citenamefont {{Santol{\'\i}k}}, \citenamefont {{Kurth}},
  \citenamefont {{Hospodarsky}}, \citenamefont {{Gurnett}},\ and\ \citenamefont
  {{Connerney}}}]{Brown2018}%
  \BibitemOpen
  \bibfield  {author} {\bibinfo {author} {\bibfnamefont {S.}~\bibnamefont
  {{Brown}}}, \bibinfo {author} {\bibfnamefont {M.}~\bibnamefont {{Janssen}}},
  \bibinfo {author} {\bibfnamefont {V.}~\bibnamefont {{Adumitroaie}}}, \bibinfo
  {author} {\bibfnamefont {S.}~\bibnamefont {{Atreya}}}, \bibinfo {author}
  {\bibfnamefont {S.}~\bibnamefont {{Bolton}}}, \bibinfo {author}
  {\bibfnamefont {S.}~\bibnamefont {{Gulkis}}}, \bibinfo {author}
  {\bibfnamefont {A.}~\bibnamefont {{Ingersoll}}}, \bibinfo {author}
  {\bibfnamefont {S.}~\bibnamefont {{Levin}}}, \bibinfo {author} {\bibfnamefont
  {C.}~\bibnamefont {{Li}}}, \bibinfo {author} {\bibfnamefont {L.}~\bibnamefont
  {{Li}}}, \bibinfo {author} {\bibfnamefont {J.}~\bibnamefont {{Lunine}}},
  \bibinfo {author} {\bibfnamefont {S.}~\bibnamefont {{Misra}}}, \bibinfo
  {author} {\bibfnamefont {G.}~\bibnamefont {{Orton}}}, \bibinfo {author}
  {\bibfnamefont {P.}~\bibnamefont {{Steffes}}}, \bibinfo {author}
  {\bibfnamefont {F.}~\bibnamefont {{Tabataba-Vakili}}}, \bibinfo {author}
  {\bibfnamefont {I.}~\bibnamefont {{Kolma{\v{s}}ov{\'a}}}}, \bibinfo {author}
  {\bibfnamefont {M.}~\bibnamefont {{Imai}}}, \bibinfo {author} {\bibfnamefont
  {O.}~\bibnamefont {{Santol{\'\i}k}}}, \bibinfo {author} {\bibfnamefont
  {W.}~\bibnamefont {{Kurth}}}, \bibinfo {author} {\bibfnamefont
  {G.}~\bibnamefont {{Hospodarsky}}}, \bibinfo {author} {\bibfnamefont
  {D.}~\bibnamefont {{Gurnett}}}, \ and\ \bibinfo {author} {\bibfnamefont
  {J.}~\bibnamefont {{Connerney}}},\ }\href {\doibase
  10.1038/s41586-018-0156-5} {\bibfield  {journal} {\bibinfo  {journal} {\nat}\
  }\textbf {\bibinfo {volume} {558}},\ \bibinfo {pages} {87} (\bibinfo {year}
  {2018})}\BibitemShut {NoStop}%
\bibitem [{\citenamefont {{Astropy Collaboration}}\ and\ \citenamefont
  {{Astropy Contributors et al.}}(2013)}]{code_Astropy2013}%
  \BibitemOpen
  \bibfield  {author} {\bibinfo {author} {\bibnamefont {{Astropy
  Collaboration}}}\ and\ \bibinfo {author} {\bibnamefont {{Astropy Contributors
  et al.}}},\ }\href {\doibase 10.1051/0004-6361/201322068} {\bibfield
  {journal} {\bibinfo  {journal} {Astronomy and Astrophysics {AAP}}\ }\textbf
  {\bibinfo {volume} {558}},\ \bibinfo {eid} {A33} (\bibinfo {year} {2013})},\
  \Eprint {http://arxiv.org/abs/1307.6212} {arXiv:1307.6212 [astro-ph.IM]}
  \BibitemShut {NoStop}%
\bibitem [{\citenamefont {{Astropy Collaboration}}\ and\ \citenamefont
  {{Astropy Contributors et al.}}(2018)}]{code_Astropy2018}%
  \BibitemOpen
  \bibfield  {author} {\bibinfo {author} {\bibnamefont {{Astropy
  Collaboration}}}\ and\ \bibinfo {author} {\bibnamefont {{Astropy Contributors
  et al.}}},\ }\href {\doibase 10.3847/1538-3881/aabc4f} {\bibfield  {journal}
  {\bibinfo  {journal} {Astronomical Journal {AJ}}\ }\textbf {\bibinfo {volume}
  {156}},\ \bibinfo {eid} {123} (\bibinfo {year} {2018})},\ \Eprint
  {http://arxiv.org/abs/1801.02634} {arXiv:1801.02634 [astro-ph.IM]}
  \BibitemShut {NoStop}%
\bibitem [{\citenamefont {{Perez}}\ and\ \citenamefont
  {{Granger}}(2007)}]{code_Perez2007}%
  \BibitemOpen
  \bibfield  {author} {\bibinfo {author} {\bibfnamefont {F.}~\bibnamefont
  {{Perez}}}\ and\ \bibinfo {author} {\bibfnamefont {B.~E.}\ \bibnamefont
  {{Granger}}},\ }\href {\doibase 10.1109/MCSE.2007.53} {\bibfield  {journal}
  {\bibinfo  {journal} {Computing in Science and Engineering}\ }\textbf
  {\bibinfo {volume} {9}},\ \bibinfo {pages} {21} (\bibinfo {year}
  {2007})}\BibitemShut {NoStop}%
\bibitem [{\citenamefont {Hunter}(2007)}]{code_Hunter2007}%
  \BibitemOpen
  \bibfield  {author} {\bibinfo {author} {\bibfnamefont {J.~D.}\ \bibnamefont
  {Hunter}},\ }\href {\doibase 10.1109/MCSE.2007.55} {\bibfield  {journal}
  {\bibinfo  {journal} {Computing in Science \& Engineering}\ }\textbf
  {\bibinfo {volume} {9}},\ \bibinfo {pages} {90} (\bibinfo {year} {2007})},\
  \Eprint
  {http://arxiv.org/abs/https://aip.scitation.org/doi/pdf/10.1109/MCSE.2007.55}
  {https://aip.scitation.org/doi/pdf/10.1109/MCSE.2007.55} \BibitemShut
  {NoStop}%
\bibitem [{\citenamefont {Walt}\ \emph {et~al.}(2011)\citenamefont {Walt},
  \citenamefont {Colbert},\ and\ \citenamefont {Varoquaux}}]{code_Walt2011}%
  \BibitemOpen
  \bibfield  {author} {\bibinfo {author} {\bibfnamefont {S.~v.~d.}\
  \bibnamefont {Walt}}, \bibinfo {author} {\bibfnamefont {S.~C.}\ \bibnamefont
  {Colbert}}, \ and\ \bibinfo {author} {\bibfnamefont {G.}~\bibnamefont
  {Varoquaux}},\ }\href {\doibase 10.1109/MCSE.2011.37} {\bibfield  {journal}
  {\bibinfo  {journal} {Computing in Science \& Engineering}\ }\textbf
  {\bibinfo {volume} {13}},\ \bibinfo {pages} {22} (\bibinfo {year} {2011})},\
  \Eprint
  {http://arxiv.org/abs/https://aip.scitation.org/doi/pdf/10.1109/MCSE.2011.37}
  {https://aip.scitation.org/doi/pdf/10.1109/MCSE.2011.37} \BibitemShut
  {NoStop}%
\bibitem [{\citenamefont {{Starkman}}(2020{\natexlab{a}})}]{code_utilipy2020}%
  \BibitemOpen
  \bibfield  {author} {\bibinfo {author} {\bibfnamefont {N.}~\bibnamefont
  {{Starkman}}},\ }\href {\doibase 10.5281/zenodo.3740639} {\bibfield
  {journal} {\bibinfo  {journal} {Zenodo}\ } (\bibinfo {year}
  {2020}{\natexlab{a}}),\ 10.5281/zenodo.3740639}\BibitemShut {NoStop}%
\bibitem [{\citenamefont
  {{Starkman}}(2020{\natexlab{b}})}]{code_starkplot2020}%
  \BibitemOpen
  \bibfield  {author} {\bibinfo {author} {\bibfnamefont {N.}~\bibnamefont
  {{Starkman}}},\ }\href {\doibase 10.5281/zenodo.3610363} {\enquote {\bibinfo
  {title} {nstarman/starkplot: v0.1.0},}\ } (\bibinfo {year}
  {2020}{\natexlab{b}})\BibitemShut {NoStop}%
\bibitem [{\citenamefont {{Starkman}}\ \emph {et~al.}(2020)\citenamefont
  {{Starkman}}, \citenamefont {{Jagjit}}, \citenamefont {{Winch}},\ and\
  \citenamefont {{Starkman}}}]{code_macrolightning2020}%
  \BibitemOpen
  \bibfield  {author} {\bibinfo {author} {\bibfnamefont {N.}~\bibnamefont
  {{Starkman}}}, \bibinfo {author} {\bibfnamefont {S.}~\bibnamefont
  {{Jagjit}}}, \bibinfo {author} {\bibfnamefont {H.}~\bibnamefont {{Winch}}}, \
  and\ \bibinfo {author} {\bibfnamefont {G.}~\bibnamefont {{Starkman}}},\
  }\href {\doibase 10.5281/zenodo.3926341} {\bibfield  {journal} {\bibinfo
  {journal} {Zenodo}\ } (\bibinfo {year} {2020}),\
  10.5281/zenodo.3926341}\BibitemShut {NoStop}%
\end{thebibliography}%


\end{document}